\documentclass[journal=jpccck,manuscript=article]{achemso}

\usepackage[T1]{fontenc}       

\usepackage{bm}
\usepackage{bbm}
\usepackage{mathrsfs}
\usepackage{color}
\usepackage{graphicx}
\usepackage{dcolumn}
\usepackage{amssymb}
\usepackage{amsmath}

\title{Dynamic characterization of cellulose nanofibrils in sheared and extended semi-dilute dispersions}

\author{Tomas Ros\'{e}n}
\email{rosen@mech.kth.se}
\affiliation{Linn\'{e} FLOW Center, KTH Mechanics, Qsquars backe 18, Royal Institute of Technology, SE-100 44 Stockholm, Sweden}
\alsoaffiliation{Wallenberg Wood Science Center, Royal Institute of Technology, SE-100 44 Stockholm, Sweden}
\alsoaffiliation{Present address: Department of Chemistry, Stony Brook University, Stony Brook, NY 11794-3400, USA}

\author{Nitesh Mittal}
\affiliation{Linn\'{e} FLOW Center, KTH Mechanics, Qsquars backe 18, Royal Institute of Technology, SE-100 44 Stockholm, Sweden}
\alsoaffiliation{Wallenberg Wood Science Center, Royal Institute of Technology, SE-100 44 Stockholm, Sweden}

\author{Stephan V. Roth}
\affiliation{DESY, Notkestrasse 85, Hamburg, Germany}
\alsoaffiliation{Department of Fibre and Polymer Technology, Teknikringen 56-58, Royal Institute of Technology, SE-100 44 Stockholm, Sweden}

\author{Peng Zhang}
\affiliation{DESY, Notkestrasse 85, Hamburg, Germany}
\alsoaffiliation{Department of Fibre and Polymer Technology, Teknikringen 56-58, Royal Institute of Technology, SE-100 44 Stockholm, Sweden}
\alsoaffiliation{Present address: INM-Leibniz Institute for New Materials, Campus D2 2, 66123 Saarbruecken, Germany}

\author{L. Daniel S\"oderberg}
\affiliation{Linn\'{e} FLOW Center, KTH Mechanics, Qsquars backe 18, Royal Institute of Technology, SE-100 44 Stockholm, Sweden}
\alsoaffiliation{Wallenberg Wood Science Center, Royal Institute of Technology, SE-100 44 Stockholm, Sweden}

\author{Fredrik Lundell}
\affiliation{Linn\'{e} FLOW Center, KTH Mechanics, Qsquars backe 18, Royal Institute of Technology, SE-100 44 Stockholm, Sweden}
\alsoaffiliation{Wallenberg Wood Science Center, Royal Institute of Technology, SE-100 44 Stockholm, Sweden}

\date{\today}

\begin{document}

\begin{abstract}
New materials made through controlled assembly of dispersed cellulose nanofibrils (CNF) has the potential to develop into biobased competitors to some of the highest performing materials today. The performance of these new cellulose materials depends on how easily CNF alignment can be controlled with hydrodynamic forces, which are always in competition with a different process driving the system towards isotropy, called \emph{rotary diffusion}. In this work, we present a flow-stop experiment using polarized optical microscopy (POM) to study the rotary diffusion of CNF dispersions in process relevant flows and concentrations. This is combined with small angle X-ray scattering (SAXS) experiments to analyze the true orientation distribution function (ODF) of the flowing fibrils. It is found that the rotary diffusion process of CNF occurs at multiple time scales, where the fastest scale seems to be dependent on the deformation history of the dispersion before the stop. At the same time, the hypothesis that rotary diffusion is dependent on the initial ODF does not hold as the same distribution can result in different diffusion time scales. The rotary diffusion is found to be faster in flows dominated by shear compared to pure extensional flows. Furthermore, the experimental setup can be used to quickly characterize the dynamic properties of flowing CNF and thus aid in determining the quality of the dispersion and its usability in material processes.
\end{abstract}

\maketitle
\section{Introduction}
Nature has a remarkable way of structuring materials in complex hierarchies resulting in properties that are hard to achieve with synthetic materials \citep{wegst2015bioinspired}. Finding a process to produce exceptional biobased materials on an industrial scale can lead to new sustainable products with potential to replace existing products relying on extensive use of fossil raw material and fresh water. Using cellulose nanofibrils (CNF) as the building block, several examples of nanostructured high performance materials including filaments \cite{Hakansson_NatComm}, films \cite{henriksson2008cellulose} and composites \cite{hamedi2014highly,mittal2017ultrastrong} have been presented in previous publications. The general strategy to improve the material properties is to tailor the structure on a nanometer-scale through manipulation of the processing conditions. 

A major obstacle of achieving full control over the material properties in the different processes, is the fact that the CNFs can vary significantly in size, shape, polydispersity and/or mechanical properties\cite{isogai2011tempo,luo2013carbon,xu2013cellulose,ma2014fabrication,sacui2014comparison,usov2015understanding,amiralian2015isolation,amiralian2015easily,chencomparative}. This will in turn depend on both the raw material source and method for producing CNF as well as biological activity and thermal degradation. Apart from individual fibrils being affected, it is also likely that their dynamic interactions will be influenced, which in turn affect the way fibrils rotate and align in process related flows. Therefore, it is important to perform dynamic characterization of the actual dispersions, especially with the same concentration, in the same types of flows as commonly used in the assembly process. The main material process of interest and the starting point for the present study is the preparation of continuous cellulose filaments through hydrodynamic alignment and assembly of CNF demonstrated by \citet{Hakansson_NatComm}.

The dynamics of dispersed non-spherical nanoparticles in flows typically can be divided into two parts: \emph{advection} (the externally forced motion of the particles due to the deformation of the surrounding medium) and \emph{diffusion} (the internally forced motion of a collection of particles towards thermodynamic equilibrium). Furthermore, the motion of the particle can be divided into translational and rotational motion. 

The translational advection of small particles such as CNF is trivial as they will follow the same motion as the fluid. Assuming an ellipsoidal shape of the non-spherical fibrils, the rotary advection due to velocity gradients of the surrounding flow is described by \citet{Jeffery}. The diffusion of the particles is typically a result of Brownian (thermal) motion of surrounding molecules, but also other interactions between particles including direct collisions, electrostatics or various complex molecular interactions. The translational diffusive motion, typically leads to a uniform spatial distribution of particles. Translational diffusion is of less interest in CNF material processes, since we assume the spatial distribution in the dispersion to always be uniform and will not be affected by the flow. The important dynamics of CNF instead is described by their rotational motion. Here, the rotary advection due to velocity gradients in the flow (such as shear and extension) will result in a non-isotropic orientational distribution. This process is always in competition with the rotary diffusion process towards an isotropic state. Since rotary diffusion of CNF is a non-trivial process and depends on many factors, it is thus important to characterize this process for different CNF samples in order to understand how the fibrils will behave in process relevant flows.

Dynamic characterization of nanoparticle dispersions to target rotary diffusion is typically done with dynamic light scattering (DLS) using coherent light, commonly achieved with a laser source. By studying the autocorrelation of the scattered light, the translational and rotary diffusion coefficients can be obtained \cite{CHENKRAUS,GLIDDEN,HAGHIGHI,BALOG,mao2017characterization}. However, due to problems of multiple scattering in the sample, the dispersions are typically required to be dilute (<0.1~vol.\%). Furthermore, the technique can only be used to study the dynamics in isotropic and thus non-flowing systems. Thus, the DLS technique is not suitable to study dynamic CNF interactions in material processes where the concentration typically is around 0.3~vol.\%. Rotary diffusion measurements also typically requires hours of measurements to produce a correlation function with acceptable statistics.

More suitable techniques for characterizing CNF at process relevant conditions are small and wide angle X-ray scattering (SAXS/WAXS). At isotropic (non-flowing) conditions, these techniques have previously been used to obtain relevant statistics of the individual particle shapes as well as the structure of fibril networks\cite{FEIGUN,CHU,chu2001small,stribeck,mao2017characterization}. SAXS has also been widely used to study the orientation of elongated nanoparticles, including CNF, in flows\cite{pfohl2007highly,trebbin2013anisotropic,silva2015,lutz2016scanning,kamada2017flow}. If X-rays would be used to study the dynamics and interactions of fibrils during de-alignment of an initially anisotropic system, long irradiation times are necessary. The damage of the radiation on the sample can thus not be neglected in such a study. 

It is possible to estimate the rotary diffusion coefficient in SAXS experiments from measurements at different positions in the flow \cite{Hakansson_SAXSALIGN}. This requires an assumption of the orientation distribution to be purely advected downstream such that different positions correspond to different instances of time during the diffusion process. This assumption might not always be applicable due to velocity gradients in the flow. 

Recently, there has been advances also in X-ray photon correlation spectroscopy (XPCS), which is equivalent to DLS but utilizing X-rays instead of visible light and thus reducing the problem of multiple scattering \cite{riese2000photon,BUSCH,FLUERASU,HOLMQVIST,WAGNER}. This technique has been used to study the advection and diffusion of dispersed nanoparticles in flowing samples. Also XPCS suffers from the downside of radiation damage on biobased samples in non-flowing systems. To avoid radiation damage and to possibly increase the scattering contrast, neutron scattering techniques have also been used on CNF systems in a similar manner as the X-ray techniques\cite{FEIGUN,mao2017characterization}. However, given the low flux of present spallation sources, such an experiment becomes very time consuming.

Another way of studying the orientation of dispersed nanofibrils is to measure the birefringence of the dispersion using polarized optical microscopy (POM) \cite{chow1985rheooptical1,chow1985rheooptical2,lim1985conformation, rosenblatt1985birefringence, WUZHOU, PHALAKORNKUL,ROGERS,liu2014orientationally}. An aligned system of CNF has a higher refractive index for light polarized perpendicular to the fibrils compared with light polarized in the parallel direction. A measurement of how the birefringence of an initially aligned system decays with time is thus a measurement of how the fibrils approach an isotropic orientation distribution, i.e. a measurement of rotary diffusion. The advantages of this method is that a laser with visible light can be used as a light source with negligible damage to the sample. Additionally, regular high-speed cameras can be used to detect the transmitted light. The disadvantage of using POM is that it only provides a relative measurement of the average alignment in the sample. Consequently, it can not provide more detailed information about the orientation distribution of fibrils.\\

Using static measurements with SAXS and POM, \citet{Hakansson_NatComm} concluded that understanding the rotational dynamics of CNF is crucial for controlling the final material performance. In a later work\cite{Hakansson_SAXSALIGN} they further indicated that the rotary diffusion process was highly dependent on the instantaneous alignment of the dispersion. In this work, we demonstrate a flow-stop experiment using POM that can be easily used and set up for quick dynamic characterization of birefringent CNF dispersions at process relevant (actually identical) concentrations and flow conditions. By combining the dynamic characterization with SAXS measurements in the stationary flow (before stop), we can also reveal details about the initial orientation distributions giving rise to the dynamical behavior.

The POM flow-stop experiment can be used as a standard characterization tool for determining the quality of CNF dispersions. The results from this characterization method can furthermore be used to properly model the dynamics of dispersed CNF in material processes, which in turn can lead to reasonable optimization strategies to improve the performance of the final material.

\section{Experimental section}
\subsection{Sample}
Cellulose nanofibrils (CNF) were prepared similarly as in the work by \citet{Hakansson_NatComm} from chemically bleached wood fibers (a mixture of 60\% Norwegian spruce and 40\% Scots pine, supplied by Domsjo AB, Sweden). The fibrils were prepared by a carboxy methylatation followed by mechanical disintegration, according to a previously reported method \citep{waagberg2008build}. Briefly, carboxy methyl groups were introduced on the surface of the fibrils (degree of substitution of 0.1) to facilitate disintegration of the fiber wall during mechanical treatment. After the chemical pre-treatment, an aqueous dispersion of the fibers was passed through a high-pressure homogenizer. The result is a CNF dispersion with a concentration of 6~g/L. 

In an additional step, the un-fibrillated and agglomerated fiber bundles were removed from the CNF dispersion as follows. The gel like dispersion was diluted to 3.3~g/L by adding deionized water, mixed thoroughly using a mechanical mixer (12000 rpm for 10 min, Ultra Turrax, IKA, Germany) and sonication (10 min, Sonics Vibracell, USA). This diluted dispersion was then centrifuged at 5000 rpm for 60 minutes, precipitate removed and the supernatant used for further studies. The dry content of the dispersion was determined by gravimetric analysis.

\subsection{Definitions of particle alignment}
\begin{figure}[t]
\includegraphics[width=0.99\textwidth]{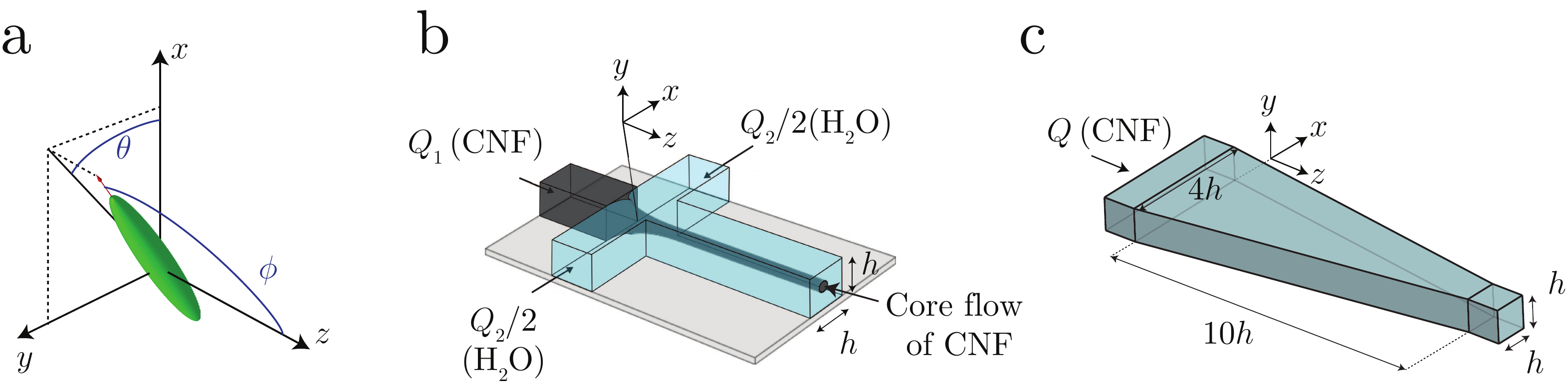}
\caption{\label{fig:FlowGeometries} (a) Definition of the particle orientation using angles $\phi$ and $\theta$; (b) illustration of the flow-focusing channel (FFC) geometry ($h=1$~mm); figure adapted from \citet{Hakansson_NatComm}; (c) the converging channel (CC) geometry.}
\end{figure}

The general flow direction in the experiment is defined to be the $z$-direction. To describe the orientation of a single fibril, we will use the polar angle $\phi$ as the angle between the fibril major axis and the flow ($z$) direction while the azimuthal angle $\theta$ is the projected angle in the plane perpendicular to the flow (see fig.~\ref{fig:FlowGeometries}a). The probability of a single fibril having a certain orientation $\phi$ and $\theta$ is given by the orientation distribution functions (ODF) $\Psi_\phi$ and $\Psi_\theta$, respectively. To describe the average alignment of the CNF in the flow direction, the order parameter $S_\phi=\frac{1}{2}\langle3\cos^2\phi-1\rangle$ will be used. The brackets in this expression denote an ensemble average over all fibrils.

\subsection{Channel geometries}
Two different channel geometries were used in this study.

The first geometry is a flow-focusing channel (FFC) as used by \citet{Hakansson_NatComm} and illustrated in fig.~\ref{fig:FlowGeometries}b. It consists of a four channel crossing, where three channels serve as inlets. The core flow of CNF (flow rate $Q_1$) in the center channel is focused by two sheath flows of water (each with flow rate $Q_2/2$) positioned opposite to each other, perpendicular to the center channel. The channels have square cross-sections of $1\times1$~mm$^2$. As characteristic length scale of the flows, we therefore choose $h=1$~mm.

The second geometry is a converging channel (CC) as illustrated in fig.~\ref{fig:FlowGeometries}c. The channel, with an initial rectangular cross-section $4\times1$~mm$^2$, is contracted linearly to a quadratic cross-section of $1\times1$~mm$^2$ over a length of 10~mm. The flow rate in the CC geometry is denoted as $Q$.

\begin{figure}[t]
\includegraphics[width=0.99\textwidth]{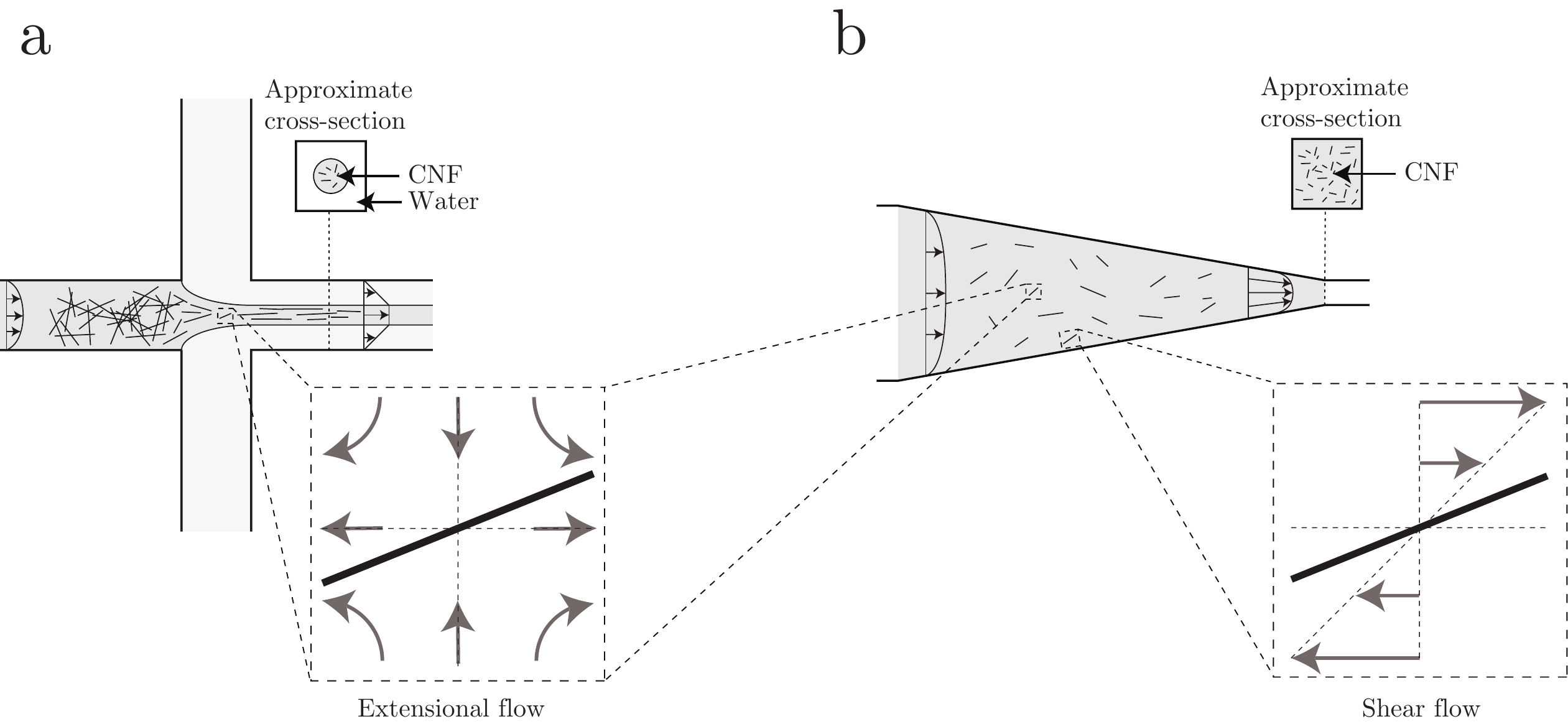}
\caption{\label{fig:Shear_and_Strain_Intro} Illustration of the different velocity gradients that the CNF dispersion is experiencing in the two channel geometries; (a) the particles on the centerline in the flow-focusing channel geometry (FFC) are experiencing a uni-axial extensional flow; (b) the particles on the centerline in the converging channel geometry (CC) are experiencing a planar extensional flow while particles close to the walls are experiencing shear flow.}
\end{figure}

In both channel geometries, the CNF dispersion is accelerated, leading to an increased alignment of the fibrils. There are however some key differences between the FFC and CC geometries that are illustrated in fig.~\ref{fig:Shear_and_Strain_Intro}a-b.

Firstly, when changing the acceleration in the FFC geometry by changing $Q_1$ and $Q_2$, also the cross-section of the core will change. Using optical techniques to characterize the alignment, e.g.~SAXS and POM, there will thus give difficulties in comparing the effect of increasing acceleration since there will also be less amount of CNF in a channel cross-section. Furthermore, at high accelerations the core flow might fluctuate spatially due to instabilities of the flow, while low accelerations can cause a transition to a flow regime where the core fluid is not detached from the walls\cite{cubaud2014regimes}.  In the CC geometry on the other hand, the acceleration can be increased with $Q$ without changing the amount of CNF in the channel cross-section. Furthermore, the flow is also more stable at high accelerations.

Secondly, in the FFC geometry, the fibrils in the core are approximately only experiencing a uni-axial extensional flow\cite{Hakansson_SAXSALIGN}. An assumption can thus be made about cylindrical symmetry around the centerline, with a uniform distribution of the angle $\theta$ ($\Psi_\theta=\text{constant}$). Furthermore, $\Psi_\phi$ can be assumed not to vary over the core cross-section. In this channel it is thus possible to easily obtain the steady state ODF $\Psi_{\phi,0}$ from SAXS experiments using the reconstruction method by \citet{Rosen_MCSAXS}. In the CC geometry, close to the walls, the particles are experiencing velocity gradients in the wall normal directions, i.e.~a shear flow in addition to the extensional flow due to the acceleration. The rotational motion of elongated particles in shear differs from that in extension; in shear, the particles are performing a flipping motion in contrast to the pure aligning occurring in extensional flow. The orientation distribution in the CC channel can thus not be assumed to be constant over the channel cross-section, as there are differences in extension/shear levels depending on the distance from the walls.

Due to these differences, the following studies will be done in the two channel geometries:
\begin{itemize}
\item In the FFC geometry, the influence of the ODF $\Psi_{\phi,0}$ on the rotary diffusion process will be studied at the same conditions as \citet{Hakansson_NatComm,Hakansson_SAXSALIGN}, i.e.~the core flow rate is $Q_1=23.4$~ml/h and the sheath flow rate is $Q_2/2=13.5$~ml/h in each of the two channels. In this geometry, the flow is assumed to only have velocity gradients in the flow direction, i.e.~a pure extensional flow.
\item In the CC geometry, the influence of acceleration on the rotary diffusion process will be studied by changing the flow rate $Q$. In this geometry, the combination of both shear and extension will be considered.
\end{itemize} 

\begin{figure}[t]
\includegraphics[width=0.99\textwidth]{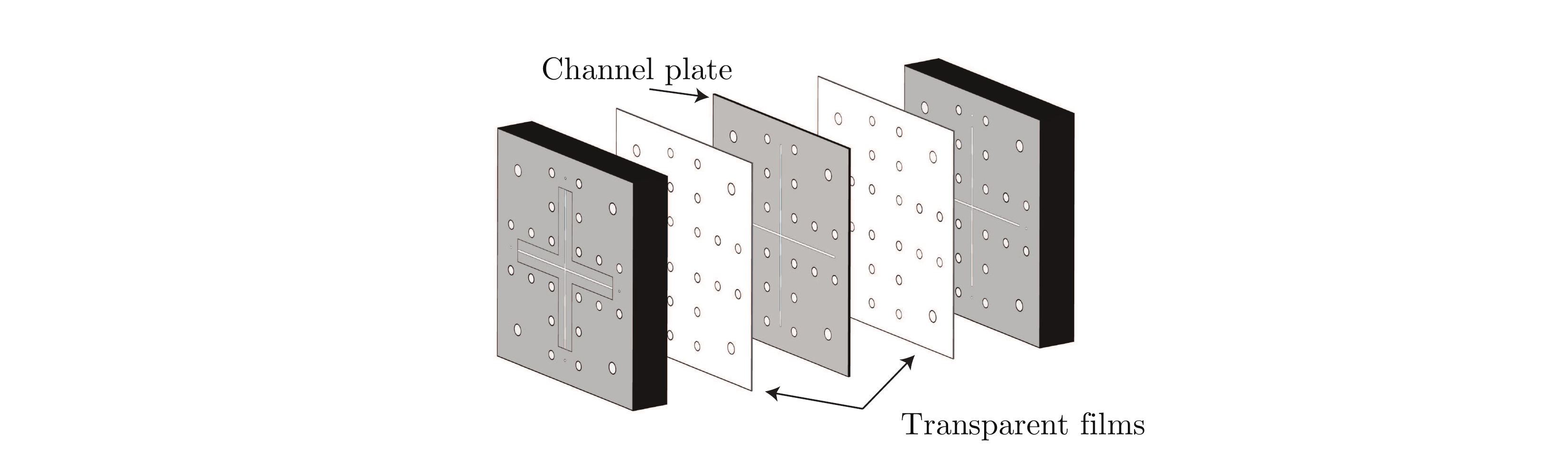}
\caption{\label{fig:Sandwich} Illustration of the assembly of the experimental flow cell; a channel plate is sandwiched between two transparent films; aluminum plates are placed outside for mechanical support.}
\end{figure}

\subsection{Flow cell}
The flow cell has a sandwich construction (fig.~\ref{fig:Sandwich}) where the channel plate is placed between two 140~$\mu$m thick transparent COC films (Tekni-plex 8007 X-04). The COC films are used instead of the PMMA-plates used by \citet{Hakansson_NatComm} since COC has beneficial optical properties and low birefringence, especially during compression, which makes it ideal for POM-experiments. For the SAXS experiments, these films are replaced by Kapton (DuPont 200HN, 50~$\mu$m thickness) to reduce damage to the films made by the high intensity X-ray beam. For mechanical support, two 10~mm aluminum plates are placed outside the films and screws are used to clamp the layers together.

\begin{figure}[t]
\includegraphics[width=0.99\textwidth]{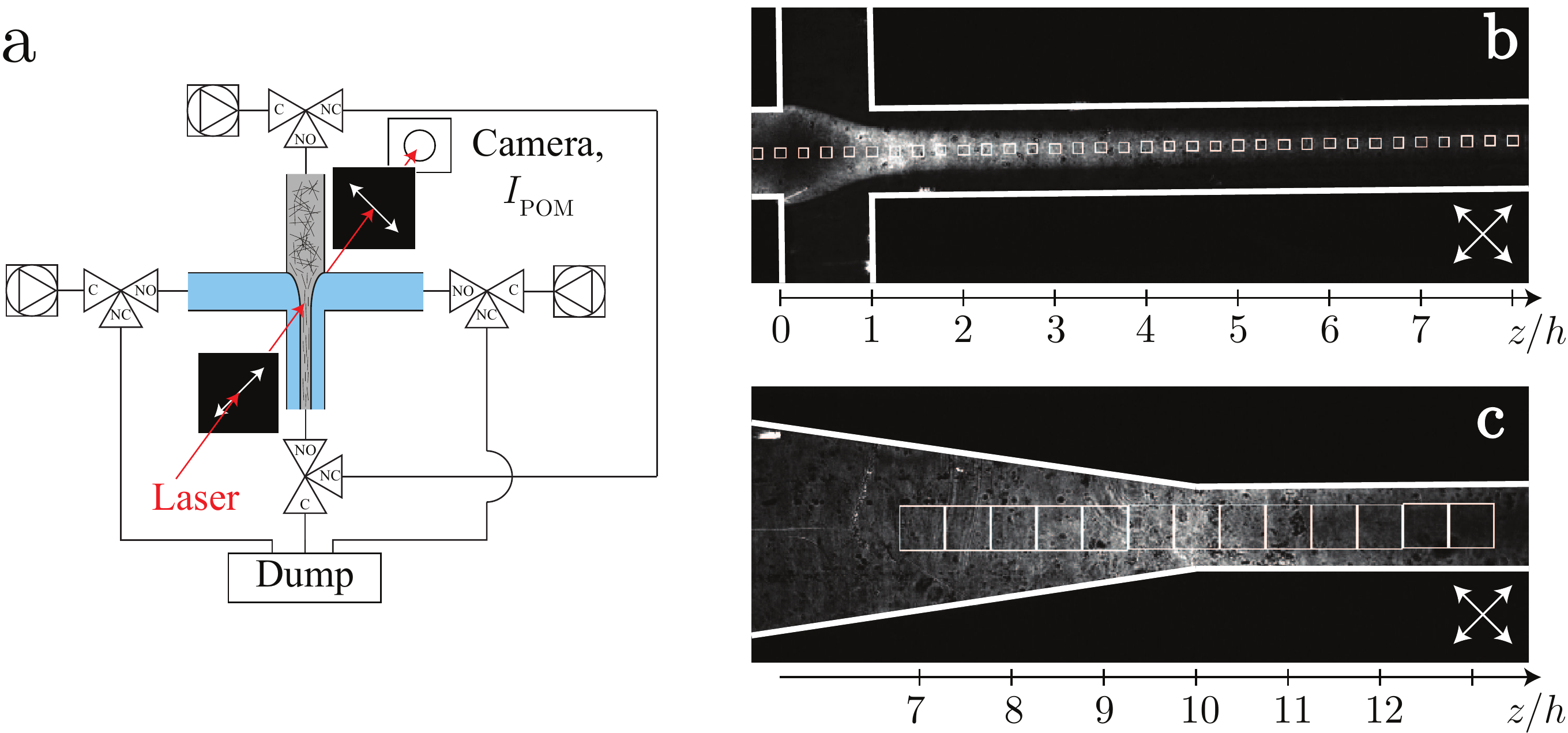}
\caption{\label{fig:POMSetup} (a) Schematic illustration of the POM flow-stop experiment; (b) example POM image during stationary flow in the FFC geometry; (c) example POM image during stationary flow in the CC geometry; white squares indicate the downstream measurement locations.}
\end{figure}

\subsection{The flow-stop experiment}
The setup of the POM flow-stop experiment is illustrated in fig.~\ref{fig:POMSetup}a. The flow cell is placed with the $z$-direction in the vertical direction between two linear polarization filters, with polarization directions $45^{\circ}$ and $-45^{\circ}$ to the flow ($z$) direction. A laser module (130~mW and wavelength of $\approx$~660~nm) is used as a light source illuminating a region of around 10~mm in diameter. The light transmitted through the setup is recorded with a Mako U-029B camera. The flow is distributed to the flow cell using syringe pumps (WPI, Al-4000) and capillary PTFE tubing. The flow can be stopped instantly in the flow cell by redirecting the flow with four synchronized three-way Takasago MTV-3SL slider valves. The camera is acquiring images with a rate of 100 frames per second during 15~s. The flow is stopped after more than 5~s of flow, which is sufficient for stationary conditions to have developed. 

In order to only observe the birefringence caused by the CNF, a background subtraction of the intensity is done using measurements with water flowing through the channel and the corrected intensity is divided with the background intensity to remove any effects of the laser light not being completely uniform over the field of view. Additionally, the intensity is further corrected separately at each measurement position by assuming that the CNF have reached isotropy after being stopped for 10~s. 

Example images from the POM experiment in both channel geometries are given in figures~\ref{fig:POMSetup}b-c, where the white squares indicate the different measurement positions. The average intensity in each of the square measurement regions $I_\text{POM}(z,t)$, where $z$ is the center of the square, is a measure of CNF alignment through the relationship\cite{vanGurp}\\
\begin{equation}
S_\phi(z,t)=C\sqrt{I_\text{POM}(z,t)},
\label{eq:OrderDependingOnIntensity}
\end{equation}
where the calibration constant $C=S_{\phi,0}(z_\text{ref.})/\sqrt{I_{\text{POM},0}(z_\text{ref.})}$ can be obtained by knowing the steady state order parameter during flow at a certain reference point $z_\text{ref.}$. Assuming a dilute dispersion of stiff rods only subject to Brownian diffusion, it is furthermore known that the order parameter decays as\cite{lim1985conformation,rosenblatt1985birefringence,Hakansson_NatComm}\\
\begin{equation}
S_\phi(z,t)=S_{\phi,0}(z)\exp(-6D_rt),
\label{eq:diluteBrownian}
\end{equation}
~\\
for an initial distribution with alignment $S_{\phi,0}(z)$. The rotary diffusion coefficient $D_r$ can thus be measured in the POM flow-stop experiment by combining the previous two equations\\
\begin{equation}
I_\text{POM}(z,t)=I_{\text{POM},0}(z)\exp(-12D_rt),
\label{eq:IntensityDr}
\end{equation}
where the steady state intensity level during flow $I_{\text{POM,0}}(z)$ is taken before stop. Further details about the post-processing of the data in the flow-stop experiment is given as supplementary information.

\begin{figure*}[t]
\includegraphics[width=0.75\textwidth]{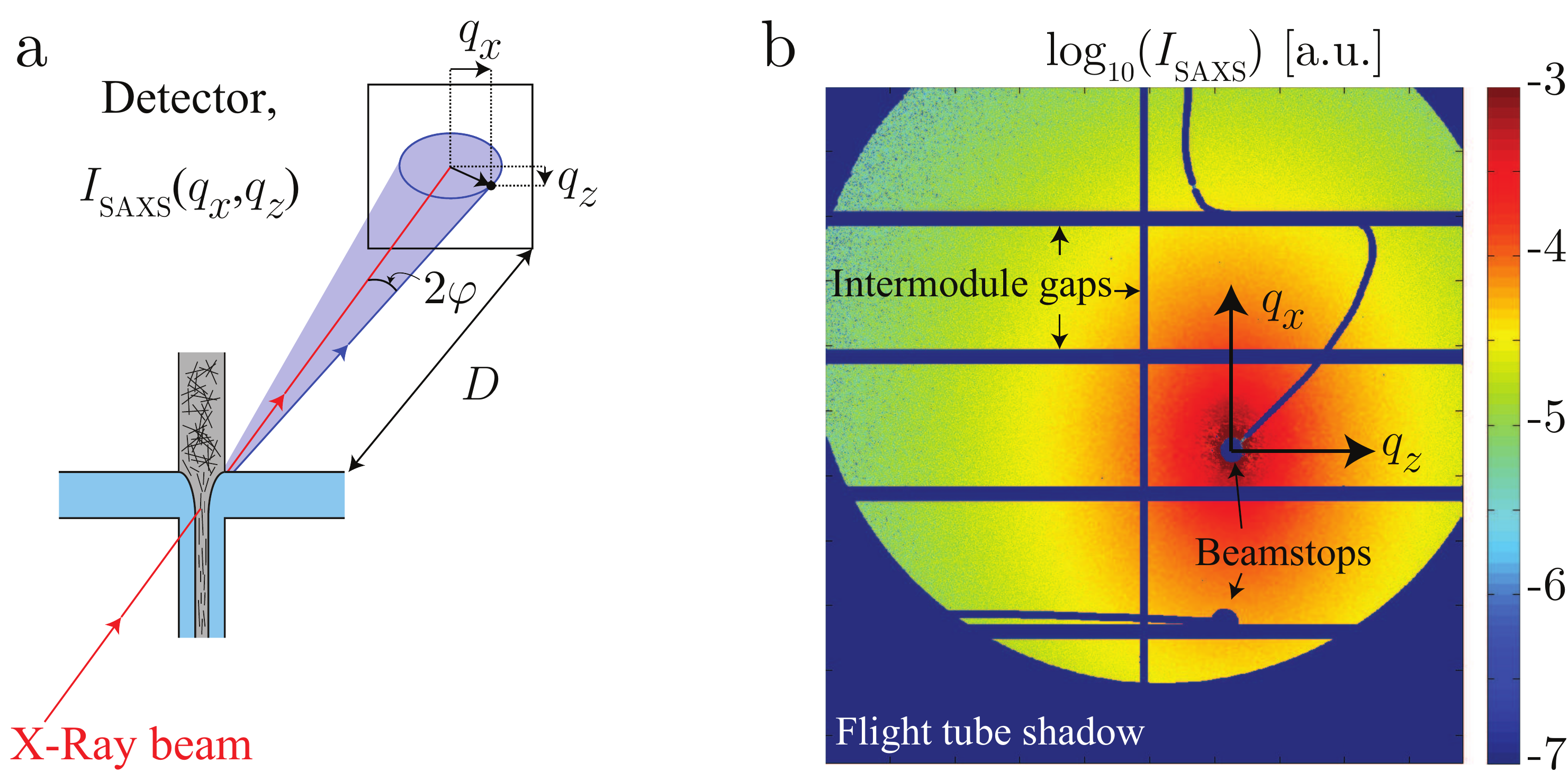}
\caption{\label{fig:Figure_SAXS} (a) Schematic illustration of the SAXS experiment; (b) example detector image from the SAXS experiment taken in the FFC geometry at $z=2h$; pixels at the intermodule gaps on the detector as well as pixels shadowed by the flight tube and beamstops contain no data and are not used in the SAXS analysis.}
\end{figure*}

\subsection{SAXS measurements}
The setup of the in situ SAXS experiments is illustrated in fig.~\ref{fig:Figure_SAXS}a. The experiments were performed at the P03 Beamline~\citep{buffet2012p03} of PETRA~III at the Deutsches Elektronen-Synchrotron (DESY) in Hamburg, Germany. The flow cell is mounted on a translation stage in front of the beam, with a detector (Pilatus 1M, Dectris, with pixel size $172\times172$~$\mu$m$^2$) placed at distance $D$ from the sample. This distance was found through calibration with a collagen sample to be $D=7.5$~m. The recorded wavelength of the X-rays is $\lambda=0.95$~\AA~and the beam area is $26\times 22$~$\mu$m$^2$. The scattered X-rays are recorded on the detector for different scattering vector lengths $q=(4\pi/\lambda)\sin\varphi$ (where $2\varphi$ is the angle between incoming and scattered radiation) during 5~s per image and in total more than 20 images per position are acquired. For the background reference scattering using only water in the channel, at least 5 images were recorded. The final analysis was then performed using an average of the images at a given position. An example of an averaged scattering image after background subtraction at position $z=2h$ is illustrated in fig.~\ref{fig:Figure_SAXS}. Note that the horizontal and vertical lines represent intermodule gaps on the detector and the other blue areas correspond to detector pixels that are shadowed by beamstops and flight tube. None of these pixels contain any scattering information and are not used in the SAXS analysis. The azimuthal anisotropy of the scattering intensity around the center of the beam represents the alignment of the CNF in the channel. The SAXS experiments where only performed on the FFC geometry and the details of how to obtain the steady state ODF $\Psi_{\phi,0}$, with the corresponding integrated order parameter $S_{\phi,0}(z)$, are provided as supplementary information.

\begin{figure}[t]
\includegraphics[width=0.75\textwidth]{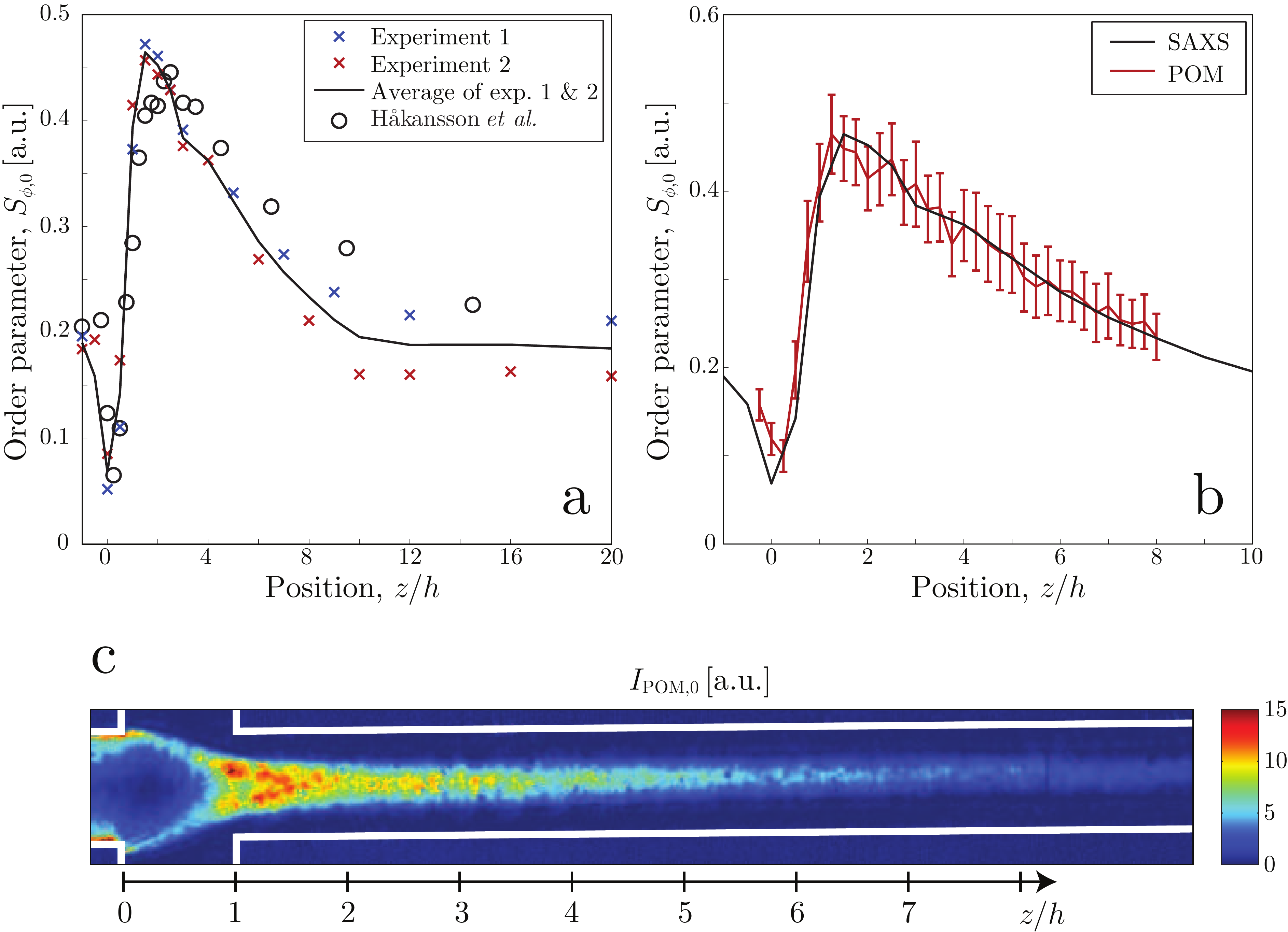}
\caption{\label{fig:FFC_SteadyState} Results of the steady state alignment in the FFC geometry; (a) order parameter $S_{\phi,0}$ versus the downstream position $z$ from the SAXS experiments and compared to the results by \citet{Hakansson_NatComm}; the black line indicates the average over two experiments; (b) results from the POM experiment compared with the SAXS experiment; (c) example of the intensity $I_{\text{POM},0}$ in an image from the POM experiment.}
\end{figure}

\section{Results: FFC geometry}
\subsection{Steady state alignment in the channel}
In fig.~\ref{fig:FFC_SteadyState}a, the results of the fibril alignment $S_{\phi,0}(z)$ are shown as function of downstream position obtained from the SAXS experiments. The average of two experiments is considered the true absolute order parameter along the channel. The results are furthermore compared with the SAXS data by \citet{Hakansson_NatComm}. Note that the values of $S_{\phi,0}(z)$ presented here are obtained by using the original data from \citet{Hakansson_NatComm} and reconstructing $\Psi_{\phi,0}$ according to the method provided by \citet{Rosen_MCSAXS}.

We find that the alignment in the channel correspond well to the previously reported ones, with a slight difference of the maximum value and the position of this peak. In the present experiments the maximum alignment is found at $z=1.5h$ with $S_{\phi,0}=0.46$, while \citet{Hakansson_NatComm} found it at $z=2.5h$ with $S_{\phi,0}= 0.45$. Although the dispersion was prepared in a similar manner here as compared to \citet{Hakansson_NatComm}, the differences in alignment could be an indication that there are differences between the two CNF dispersions.

The POM flow-stop experiments were evaluated at the downstream positions as indicated in fig.~\ref{fig:POMSetup}b. Using the position of maximum alignment as our reference position, i.e.~$z_\text{ref.}=1.5h$, and using the maximum order parameter from the SAXS experiments, we can find the calibration constant $C$ in eq.~(\ref{eq:OrderDependingOnIntensity}) and thus convert the measured POM intensity $I_\text{POM}$ to an order parameter $S_\phi$.

The resulting order parameter $S_{\phi,0}(z)$ obtained from the POM experiments is presented in fig.~\ref{fig:FFC_SteadyState}b, where the error-bars correspond to the standard deviation over 18 separate experiments. Compared to the SAXS data, which also has some scatter between the experiments, we find very good agreement using POM. It is thus sufficient to calibrate the POM experiment with only one reference position in the channel.
 
\begin{figure}[t]
\includegraphics[width=0.75\textwidth]{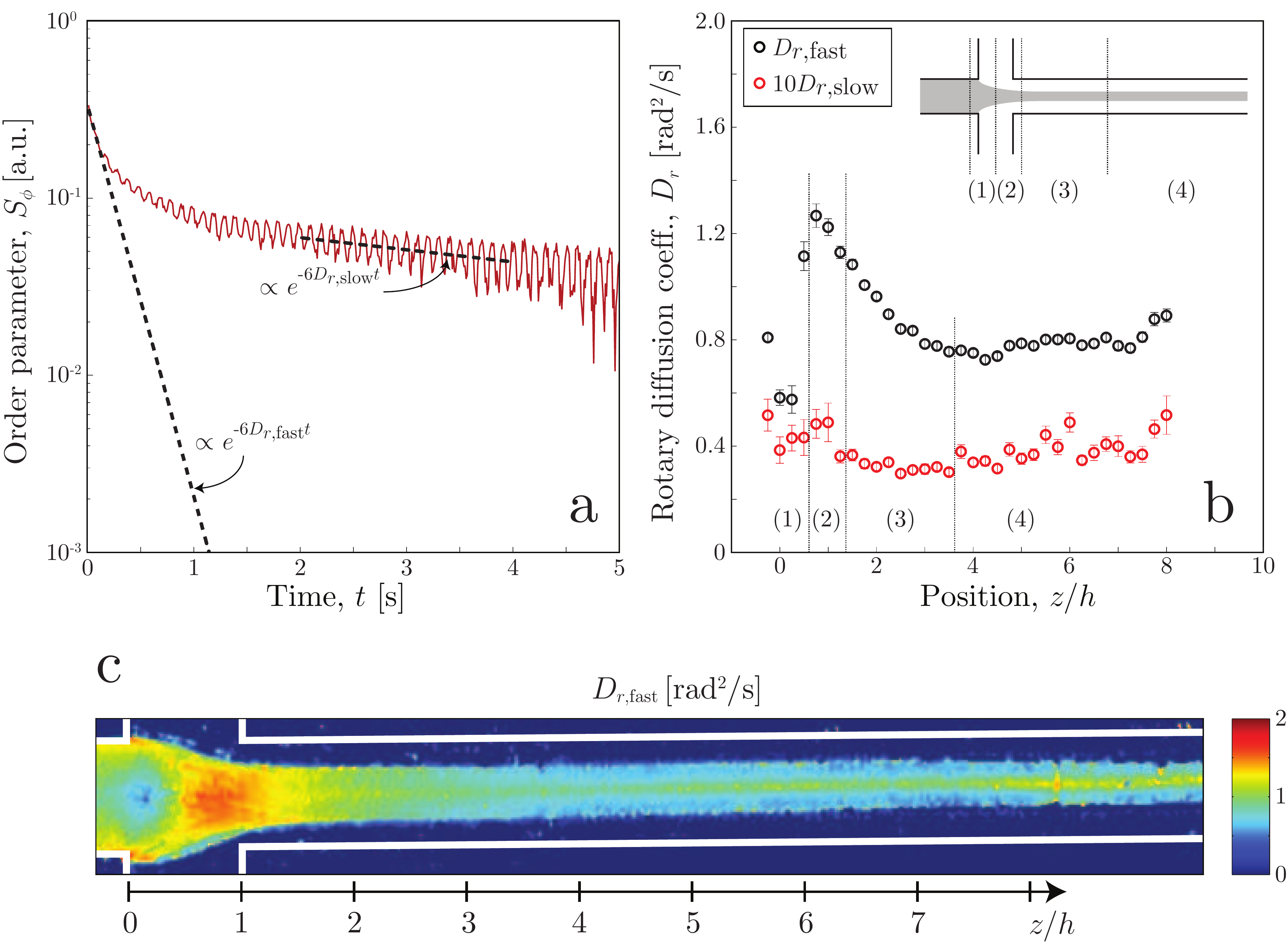}
\caption{\label{fig:FFC_Dynamic} Results from the dynamic characterization in the FFC geometry; (a) example of how the order parameter $S_\phi$ decays after stop at $z=2h$; the initial decay is proportional to $\exp(-6D_{r,\text{fast}}t)$ and the decay at longer times is proportional to $\exp(-6D_{r,\text{slow}}t)$; (b) the rotary diffusion coeffiecients $D_{r,\text{fast}}$ and $D_{r,\text{slow}}$ as functions of downstream position $z$; (c) image of how the fast rotary diffusion coefficient $D_{r,\text{fast}}$ varies along the channel.}
\end{figure}

\subsection{Dynamic characterization}
The decay of the order parameter $S_\phi$ at $z=2h$ as function of time $t$ after stop is illustrated in fig.~\ref{fig:FFC_Dynamic}a. Note that the light source used in the experiment is not perfectly stable and typically has small intensity fluctuations around its mean intensity. The oscillations seen in fig.~\ref{fig:FFC_Dynamic}a should thus not be considered to be caused by oscillations of birefringence. The order parameter $S_\phi(t)$ is plotted on a logarithmic scale. Therefore, if the dispersion would be dilute and only affected by Brownian motion, we would expect a straight line according to eq.~(\ref{eq:diluteBrownian}). However, this is clearly not the case and the decay seems to occur on multiple time scales where the alignment is initially decaying very rapidly, while at longer times the decay is slower. To analyze this further, we define two rotary diffusion coefficients at short and long times. The data at $0$~s~$<t<0.1$~s is fitted to an exponential function proportional to $\exp(-6D_{r,\text{fast}}t)$ and the data between at $2$~s~$<t<4$~s is fitted to a function proportional to $\exp(-6D_{r,\text{slow}}t)$.

Using the same procedure at different points along the channel, we find $D_{r,\text{fast}}$, $D_{r,\text{slow}}$ as functions of position $z$ illustrated in fig.~\ref{fig:FFC_Dynamic}b. Firstly, it seems that the slow diffusion is fairly constant around $D_{r,\text{slow}}\approx 0.05$~rad$^2$s$^{-1}$ and only weakly affected by the position. Secondly, we find that $D_{r,\text{fast}}$ is different depending on downstream position. Initially before the focusing region, the value is around $D_{r,\text{fast}}=0.8$~rad$^2$s$^{-1}$. However, in the acceleration zone between $0<z<h$, this effect quickly becomes stronger with a maximum value of $D_{r,\text{fast}}=1.27$~rad$^2$s$^{-1}$ at $z=0.75h$. Even though the alignment of the CNF increases between $h<z<1.5h$ (seen in fig.~\ref{fig:FFC_SteadyState}), the fast diffusion coefficient decays. Further downstream in the channel there seems to be an equilibrium value similar to the one found upstream before the focusing region. In fig.~\ref{fig:FFC_Dynamic}c, each pixel is analyzed separately to see the spatial distribution of the fast decay in the channel. In the acceleration zone, the coefficient $D_{r,\text{fast}}$ seems to be the same over the whole cross section. However, further downstream the fastest decay rates are measured close to the centerline.

\begin{figure}[t]
\includegraphics[width=0.99\textwidth]{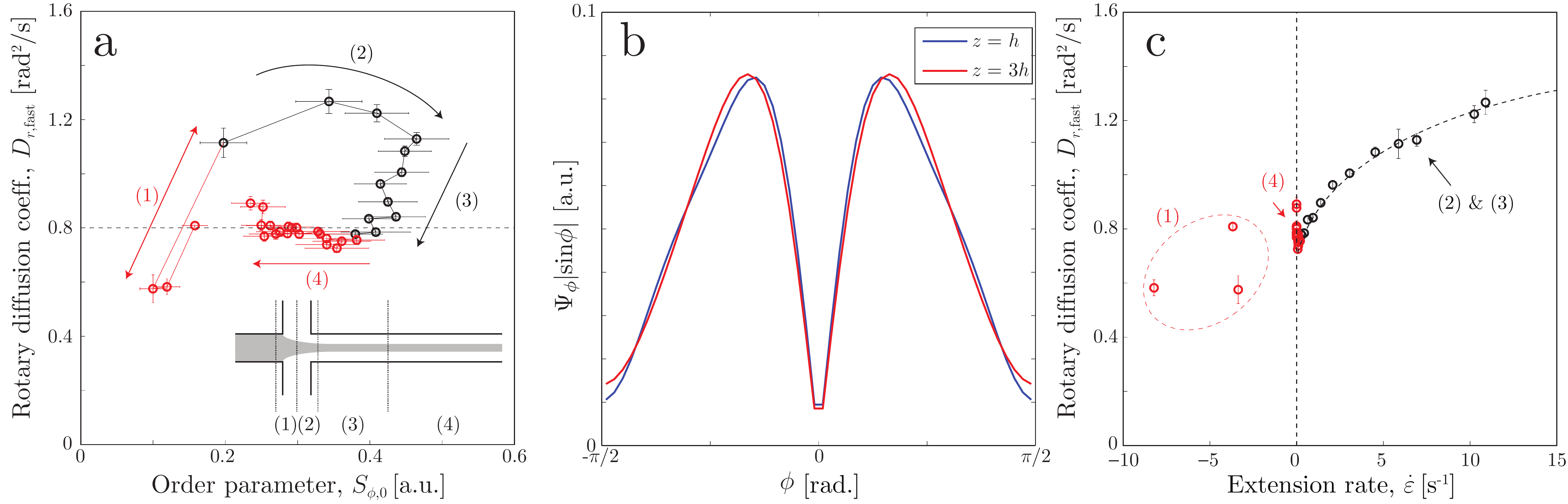}
\caption{\label{fig:FFC_Dynamic2} (a) Fast rotary diffusion coefficient $D_{r,\text{fast}}$ as function of initial steady state order parameter $S_{\phi,0}$; (b) reconstructed steady state ODF $\Psi_{\phi,0}$ at position $z=h$ and $z=3h$; the ODF is multiplied with $|\sin\phi|$ to illustrate the true probability density on the unit sphere; (c) $D_{r,\text{fast}}$ as function of the local extension rate $\dot{\varepsilon}$; the numbers (1)-(4) correspond to the same regions as illustrated in fig.~\ref{fig:FFC_Dynamic}b.}
\end{figure}

The assumption that $D_r$ only depends on the order parameter $S_{\phi,0}$ seems not to hold and it becomes even more evident when plotting the two parameters against each other in fig.~\ref{fig:FFC_Dynamic2}a. The results in the present work indicate a clear hysteresis, where the same order parameter $S_{\phi,0}$ can yield two different values of $D_r$ depending on if the measurement is upstream or downstream of the focusing region. For example at $z=h$ and $z=3h$ the order parameters are $S_{\phi,0}=0.38$~and~$0.39$, respectively. However, even though the alignment is almost the same at the two positions, there is a big difference of $D_{r,\text{fast}}$ ($1.22$~rad$^2$s$^{-1}$ and $0.78$~rad$^2$s$^{-1}$, respectively). Since the order parameter $S_{\phi,0}$ still only is an integrated value of the actual ODF, the difference could possibly be accredited to a difference in the ODFs at the two positions. To check this assumption, the two steady state ODFs $\Psi_{\phi,0}$ obtained from the SAXS experiments at these positions are illustrated in fig.~\ref{fig:FFC_Dynamic2}b. From this it is reasonable to say that the ODFs are more or less the same for a given order parameter $S_{\phi,0}$ regardless if the alignment is increased due to the acceleration or decreased due to rotary diffusion. Thus, the difference in $D_{r,\text{fast}}$ between $z=h$ and $z=3h$ does not depend on the ODF. 

By using the centerline velocity in the channel for the same flow conditions from \citet{Hakansson_SAXSALIGN}, it is possible to plot $D_{r,\text{fast}}$ as function of the local extension rate $\dot{\varepsilon}$, which is illustrated in fig.~\ref{fig:FFC_Dynamic2}c. Apart from the deceleration region with negative extension rates, it is evident that there is a clear trend where $D_{r,\text{fast}}$ is an increasing function of $\dot{\varepsilon}$. There is thus no hysteresis in figure \ref{fig:FFC_Dynamic2}c, and $D_{r,\text{fast}}$ follows the same function when increasing or decreasing the extension rate. The reason for the hysteresis in fig.~\ref{fig:FFC_Dynamic2}a is merely a consequence of the fact that $S_{\phi,0}$ increases as long as the average hydrodynamic forcing on the fibrils is overcoming rotary diffusion. Positions with the same extensional rate can therefore have different order.

The conclusion from the experiments in the FFC geometry is thus that the rotary diffusion process of CNF is mainly depending on the local velocity gradient that the fibrils experience before stopping the flow. The velocity gradient causes alignment of fibrils, but the ODF itself of the fibrils seems not to explain the variations in rotary diffusion (\emph{cf.} figure  \ref{fig:FFC_Dynamic2}b).

\section{Results: CC geometry}
\subsection{Steady state alignment in the channel}
The experiments in the converging channel (CC) geometry were performed at the different downstream positions marked with white squares in fig.~\ref{fig:POMSetup}c. In this section, the results are mean values and standard deviations over 7 experiments for each flow rate $Q$.

\begin{figure}[t]
\includegraphics[width=0.75\textwidth]{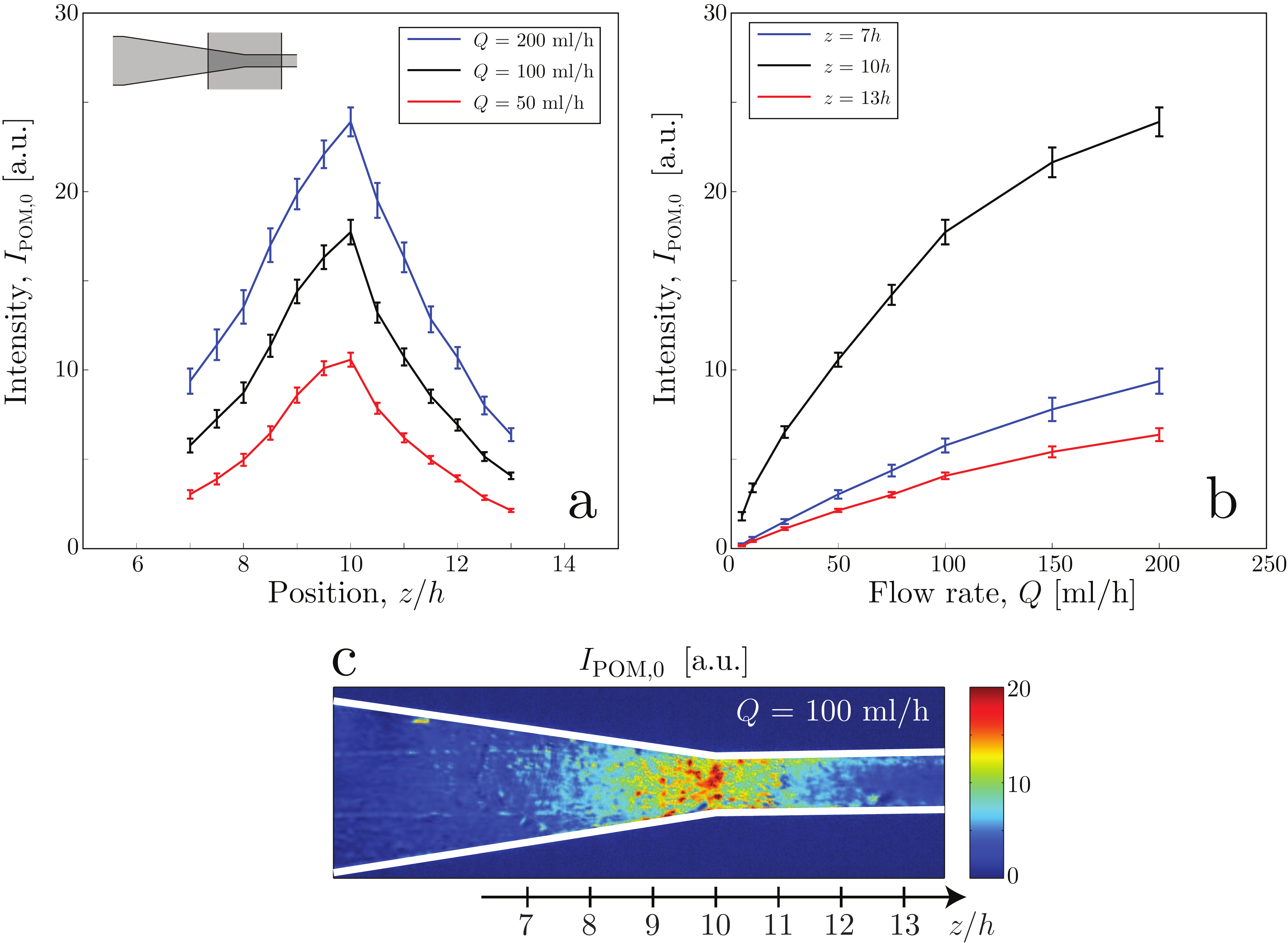}
\caption{\label{fig:CC_SteadyState} Results of the steady state alignment in the CC geometry; (a) intensity $I_{\text{POM},0}$ versus the downstream position $z$ for three different flow rates $Q$; (b)  intensity $I_{\text{POM},0}$ versus the flow rate $Q$ at three different positions; (c) example of the intensity $I_{\text{POM},0}$ in an image from the POM experiment at $Q=100$~ml/h.}
\end{figure}

The average velocity $w$ and extension rate $\dot{\varepsilon}$ in a cross section of the channel can be estimated from the geometry (neglecting the effect of the shear towards the wall):\\
\begin{equation}
w(z)=\frac{Q}{A(z)},
\end{equation}
\begin{equation}
A(z)=
\begin{cases}
4h^2,\qquad z<0\\
\frac{1}{10}(40h^2-3hz),\qquad 0<z<10h\\
h^2,\qquad z>10h
\end{cases}
\end{equation}
\begin{equation}
\dot{\varepsilon}(z)=\frac{dw}{dz}=
\begin{cases}
\frac{3Qh}{10A(z)^2},\qquad 0<z<10h\\
0,\qquad \text{otherwise}\
\label{eq:CCStrainRate}
\end{cases}
\end{equation}
~\\
At a given flow rate $Q$, there is thus an increase in extension rate $\dot{\varepsilon}$, with a maximum at $z=10h$. Since $A(z)$ is constant at $z>10h$, the extension rate rapidly decays to zero at the end of the contraction. 

In fig.~\ref{fig:CC_SteadyState}a, the steady state alignment of the fibrils in the channel is indicated by the intensity $I_{\text{POM},0}$ at different flow rates $Q$. As expected, the highest alignment is found at the end of the contraction before the rapid decay of the extension rate. Furthermore, by increasing the flow rate $Q$ at this position, the alignment increases even more as shown in fig.~\ref{fig:CC_SteadyState}b. 

\begin{figure}[t]
\includegraphics[width=0.75\textwidth]{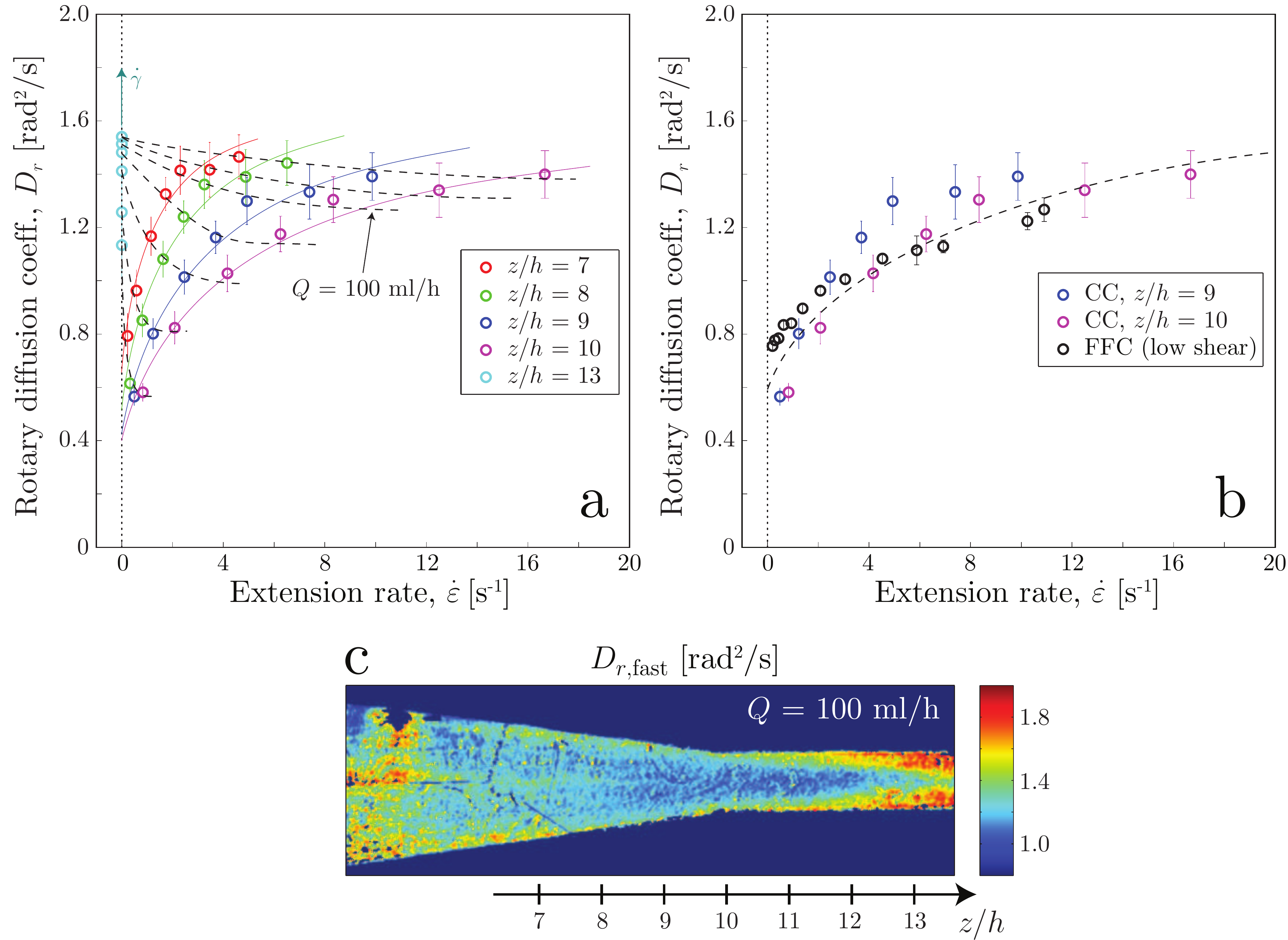}
\caption{\label{fig:CC_Dynamic} Results from the dynamic characterization in the CC geometry; (a) fast rotary diffusion coefficient $D_{r,\text{fast}}$ as function of average extension rate $\dot{\varepsilon}$; solid curves indicate approximate trends for constant downstream position $z$; dashed curves indicate approximate trends for constant flow rate $Q$; (b) comparison with the results from the FFC geometry (with low shear); the dashed curve indicates the approximate trend of $D_{r,\text{fast}}$ versus $\dot{\varepsilon}$ when shear is negligible; (c) image of how the fast rotary diffusion coefficient $D_{r,\text{fast}}$ varies along the channel at $Q=100$~ml/h.}
\end{figure}

\subsection{Dynamic characterization}
The flow stop experiment was performed in the same manner as for the FFC geometry. However, the intensity $I_{\text{POM}}$ was directly used to find $D_{r,\text{fast}}$ (with 0~s<t<0.1~s) and $D_{r,\text{slow}}$ (at 2~s<t<4~s)  according to eq.~(\ref{eq:IntensityDr}) without converting to an order parameter.

Just as for the FFC geometry, the decay of alignment after stop consists typically of a fast and a slow part described by rotary diffusion coefficients $D_{r,\text{fast}}$ and $D_{r,\text{slow}}$. The slow part however was more difficult to measure in regions of low initial alignment, as the intensity quickly comes too close to the background level. Instead, we focus the results on the fast diffusion coefficient $D_{r,\text{fast}}$, which also here varies significantly depending on downstream position. In fig.~\ref{fig:CC_Dynamic}c, the fast diffusion coefficient is obtained at $Q=100$~ml/h for each pixel and it is clear that it does not correlate well with the average alignment at the same position in figure \ref{fig:CC_SteadyState}c. Here, $D_{r,\text{fast}}$ actually has a local minimum at the position of highest alignment and highest extension rate ($z=10h$) for a given flow rate $Q$. This seems to be completely different compared to the results for the FFC geometry. Using eq.~(\ref{eq:CCStrainRate}), the coefficient $D_{r,\text{fast}}$ is plotted versus the extension rate $\dot{\varepsilon}$ for different flow rates $Q$ and positions $z$ in fig.~\ref{fig:CC_Dynamic}a. The results at the end of the contraction are then compared with the results from the FFC geometry in fig.~\ref{fig:CC_Dynamic}b. Given the same value of $Q$, it is clear that $D_{r,\text{fast}}$ decreases with increasing $\dot{\varepsilon}$. However, for a given position $z$, the coefficient $D_{r,\text{fast}}$ actually increases with $\dot{\varepsilon}$. At the end of the contraction, at $z=10h$ (purple markers in fig.~\ref{fig:CC_Dynamic}b), the apparent relationship between $D_{r,\text{fast}}$ and $\dot{\varepsilon}$ seems to be similar to the one found in the FFC geometry (black markers in fig.~\ref{fig:CC_Dynamic}b). 

The reason for the difference between different positions could be related to the average velocity gradients in the cross-stream directions, i.e.~the shear rate $\dot{\gamma}$. The shear rate will also naturally increase when moving through the contraction. After the contraction at $z>10h$, the shear rate remains high, while the extension rapidly vanishes. Increasing the flow rate $Q$ at $z=13h$, the shear rate increases along with $D_{r,\text{fast}}$, which is visible in fig.~\ref{fig:CC_Dynamic}a. From these results, it seems that $D_{r,\text{fast}}$ is even more affected by shear than by extension. At a given $Q$, when moving through the contraction, the influence from extension becomes higher than from shear and by the end of the contraction, the value of $D_{r,\text{fast}}$ is almost what is expected when shear is negligible, i.e.~close to the value obtained in the FFC geometry of around $D_{r,\text{fast}}\approx 1.3$~rad$^2$s$^{-1}$. Assuming a circular cross-section with the same area as the actual quadratic channel, the average shear rate $\dot{\gamma}$ in the fully developed flow downstream of the contraction can be estimated to be (see supplementary information for details):\\
\begin{equation}
\dot{\gamma}=\frac{8Q\sqrt{\pi}}{3h^3}.
\end{equation}
~\\
An estimation of the average shear rate at $Q=200$~ml/h would thus be $\dot{\gamma}\approx 263$~s$^{-1}$, i.e. almost 15 times higher than the maximum extension rate at the same flow rate. Even though the average shear rate at $z=10h$ should be of this order of magnitude as well, the regions of high shear typically concentrated close to the walls due to the contraction. Thus, there should be lower shear in the rest of the cross section.

In conclusion, the CC geometry can be used to predict the dynamics of CNF both in pure shear and pure extension. The measured dynamics at the end of the contraction can be assumed to not be influenced by shear and will consequently give a prediction of how the dispersed CNF will behave in the FFC geometry. Measuring further downstream of the contraction at a point where the flow is fully developed, the dynamics can similarly be assumed to not be influenced by extension but instead controlled by shear.

\section{Discussion}
Understanding the true nature of the rotary diffusion process of CNF is not trivial and requires full knowledge of all the interactions within the dispersion. Since the concentration of the dispersion studied is in the semi-dilute regime and the fibrils are electrostatically charged, the fibril-fibril interactions cannot be neglected. At the same time, the fibrils in the dispersion are typically varying in shape and size, while at the same time their mechanical properties (flexibility, stiffness etc.) are unknown in the dispersed state. We focus the discussion here on the possible causes for the experimental observations including: (1) factors influencing the order of magnitude of $D_r$; (2) factors influencing the multiple time scales in the decay of alignment; (3) the main differences between a sheared and an extended dispersion. 

\subsection{Theoretical considerations in an ideal system}
Assuming only Brownian motion in a dilute dispersion of non-interacting stiff rods, the rotary diffusion coefficient is found through\cite{doi1986theory}:\\
\begin{equation}
D_{r,\text{dilute}}=\frac{3k_bT(2\ln(2r_p)-1)}{16\pi\eta_s a^3},
\label{eq:Dr1}
\end{equation}
where $k_b$ is the Boltzmann constant, $T$ is the temperature, $r_p$ is the particle aspect ratio, $\eta_s$ is the solvent viscosity and $a$ is the half length of the fibril. Using the expected values of $k_b=1.38\times10^{-23}$~m$^2$kg~s$^{-2}$~K$^{-1}$, $T=293$~K, $r_p=100$, $\eta_s=1$~mPa~s and $l/2=a=1$~$\mu$m, the coefficient is found to be $D_{r,\text{dilute}}\approx 2$~rad$^2$s$^{-1}$. In a semi-dilute system of isotropic particles, the particle-particle interactions increase and the individual fibrils can not move as freely as in the dilute case. This hindrance due to the concentration was described with the tube-model by \citet{doi1978dynamics} leading to the expression:\\
\begin{equation}
D_r=\beta D_{r,\text{dilute}} (nl^3)^{-2},
\label{eq:Dr2}
\end{equation}
~\\
where $nl^3$ is the amount of fibrils inside a cubic volume with the same side as the fibril length. The constant $\beta$ depends on the nature of the particle-particle interactions and ranges between $1$ and $10^3$. The value must be determined experimentally. For the concentration of fibrils here, assuming mass fraction to be roughly equal to the volumetric fraction, a value of $nl^3\approx 10$ is obtained. Depending on $\beta$, it is thus expected to find the rotary diffusion coefficient in the interval $D_r\in[0.02,20]$~rad$^2$s$^{-1}$, which is also the range within which we find the present experimental results. Equations (\ref{eq:Dr1}) and (\ref{eq:Dr2}) also highlight the effects of concentration (quadratic dependence) and fibril lengths (cubic dependence), which dramatically influence the value of $D_r$. 

\subsection{Reasons for the multiple time scales}
A major factor for the multi-time-scale decay of alignment and the dependence on the velocity gradients could be accredited to the fact the CNF dispersion usually is polydisperse and consists of fibrils with various lengths. This conclusion were drawn in several similar POM flow-stop experiments using a Couette apparatus \cite{chow1985rheooptical1,chow1985rheooptical2,ROGERS}. The slow decay will thus correspond to the longest fibrils and the fast decay will correspond to the shortest fibrils that are affected by the velocity gradients. Since the stronger gradient will align shorter fibrils, the fastest decay rate also increases. However, there might be several other factors that also come into play that could also lead to a multiple time scale relaxation process.

With the same tube-model that lead to eq.~(\ref{eq:Dr2}), \citet{doi1986theory} also discussed the influence of particle alignment on the rotary diffusion coefficient. The argument was that there was less hindrance from nearby particles in an aligned system, and that $D_r$ would be an increasing function of alignment. \citet{Hakansson_SAXSALIGN} argued that the electrostatic torque imposed by nearby charged fibrils would also increase as the system aligns, which would result in a $D_r$ that increases with alignment. An additional hypothesis is that the fibrils could possibly get stretched from an initial non-straight equilibrium shape, and that the elasticity of the individual fibrils causes them to relax to the equilibrium shape; a process which would present itself as a $D_r$ that increases with alignment. Hypothetical inter-fibril bonds with some torsional stiffness\cite{stimatze2016torsional} created at these concentrations could potentially also act in a similar manner. Given any of these hypothetical effects, it is not surprising that the decay of the order parameter after stop is larger initially when the alignment also is higher. Interestingly, even though all of these effects might play a role in the rotary diffusion process, the fastest time scales are not observed at the positions of high initial alignment, but at positions of high initial shear.

However, it is believed that there is no single explanation of the measured rotary diffusion process. Factors like the polydispersity, concentration, electrostatics, individual fibril properties and other complex molecular interactions will all contribute. In order to distinguish the respective effects, new experiments are needed with well defined dispersions.

\begin{figure}[t]
\includegraphics[width=0.75\textwidth]{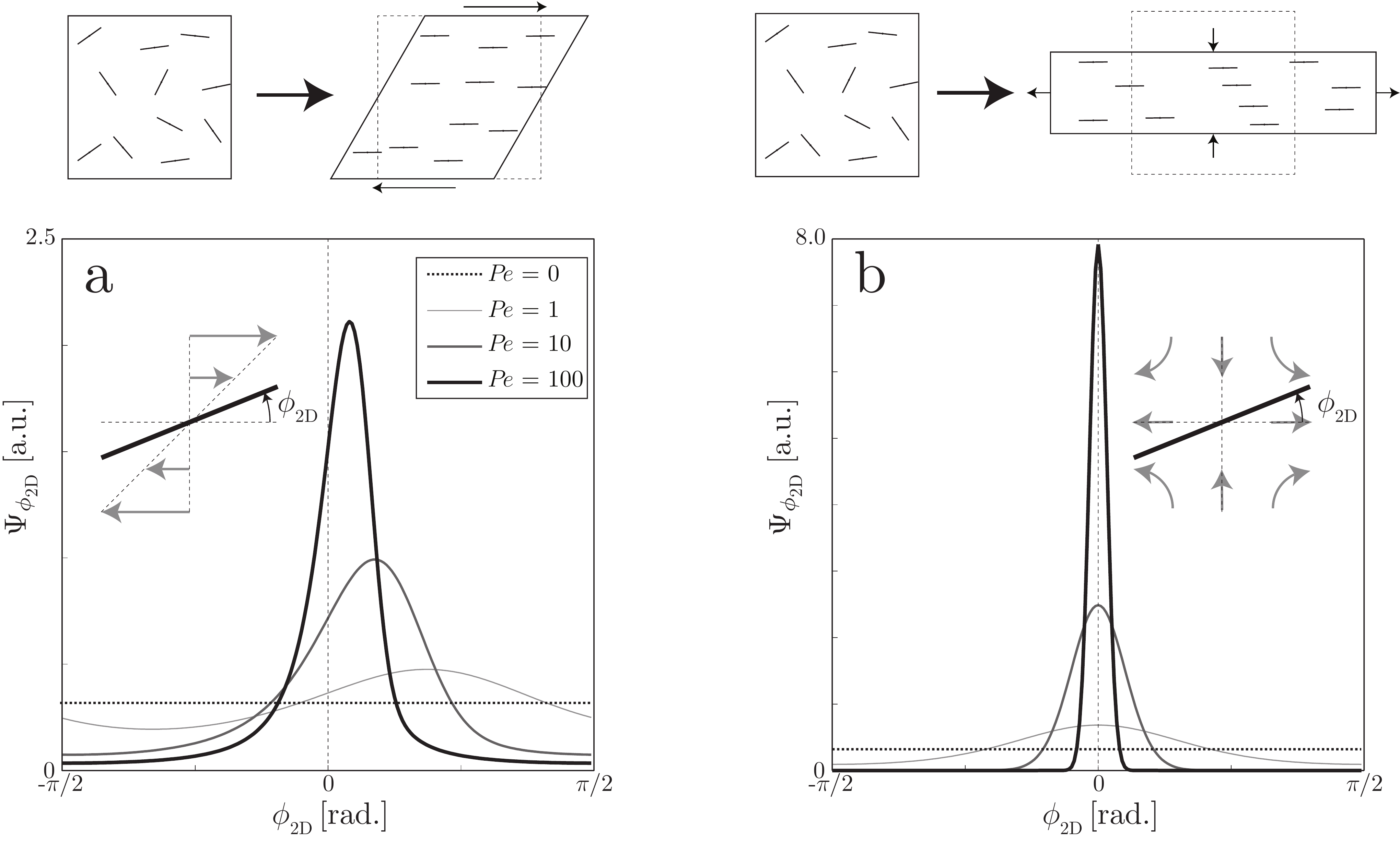}
\caption{\label{fig:Strain_vs_shear} Illustration of the difference between sheared and extensional flow in a system of CNF; the top figures show schematically how the operations affect the spatial distribution of particles; (a) the ODF $\Psi_{\phi_\text{2D}}$ for a \emph{sheared} dilute dispersion of elongated particles at different $Pe=\dot{\gamma}/D_r$; (b) the ODF $\Psi_{\phi_\text{2D}}i$ for an \emph{extended} dispersion at different $Pe=\dot{\varepsilon}/D_r$.}
\end{figure}

\subsection{Differences between shear flow and extensional flow}

The results in this work indicate that there might be significant differences in CNF dynamics between a sheared dispersion compared to the extended dispersion. The first observation in the CC geometry is that the extensional flow is much more effective in aligning the system and the system quickly de-aligns as soon as the acceleration vanishes even though there is strong shear in the channel. The second observation is that we have a significantly higher value of $D_{r,\text{fast}}$ when shear is dominating the flow downstream of the contraction. 

To illustrate the difference between the sheared and extended system, two-dimensional simulations using the Smoluchowski equation are performed. A system of non-interacting Brownian fibrils are randomly oriented in a plane with angle $\phi_\text{2D}$ to the shearing/stretching direction. More details about the method is described as supplementary information. The system is either sheared or extended (planar extension) until reaching steady state at constant P\'{e}clet number $Pe$, which is defined as $Pe=\dot{\gamma}/D_r$ (shear flow) or $Pe=\dot{\varepsilon}/D_r$ (extensional flow). 

The results of the steady state ODF $\Psi_{\phi_\text{2D}}$ are shown in fig.~\ref{fig:Strain_vs_shear}. Given the same $Pe$, the system is much less aligned in the sheared case and the ODF is heavily skewed towards positive values of $\phi_\text{2D}$. The reason is that the fibril orientation at $\phi_\text{2D}=0$ is marginally stable in the shear flow. If Brownian rotary diffusion pushes the fibril to positive values of $\phi_\text{2D}$ it will return to the original orientation. However, a push towards negative values of $\phi_\text{2D}$, will lead to the fibril \emph{flipping} half a revolution. The result is that the most probable orientation actually is larger than zero. In the extensional flow, the orientation  $\phi_\text{2D}=0$ is asymptotically stable and Brownian rotary diffusion will not lead to any flipping of the fibrils. The result is that the fibrils are aligning more effectively in the extensional flow as well as the ODF being symmetric around $\phi_\text{2D}=0$. 

The probable skewing of the ODF in the shear layer in the CC geometry means that we do not correctly measure the maximum alignment since the polarization filters are tilted 45$^{\circ}$ to the flow direction. Additional experiments using different angles of the polarization filters could possibly reveal if the ODF indeed is skewed close to the visible walls.

The reason for the faster dynamics in the sheared dispersion of CNF could be explained by the spatial distribution of the aligned fibrils. Both shear and extension will cause the fibrils to preferentially align in the flow direction. However, in the extended system, the transverse distance between fibrils is actually closer than in the sheared system as illustrated in fig.~\ref{fig:Strain_vs_shear} and fibrils located transversely do not have any relative motion. This could cause potential short-range interactions, for example caused by mechanical friction or chemical bonds, to occur more frequently and thus result in a slower rotary diffusion process. Some resistive short-range interactions could thus possibly explain the difference in behavior between a sheared and a extended dispersion. 

The idea of a rotary diffusion coefficient that is dependent on the strain rate tensor magnitude is not new and has been widely addressed in semi-dilute suspension of (non-Brownian) macroscopic fibers \cite{folgar1984orientation,koch1995model,krochak2008orientation}, where the rotary diffusion process itself is a result of short-range interactions between fibers. In these cases, it is argued that these the rate of these interactions, such as direct fiber-fiber collisions, increases with the magnitude of the strain rate tensor and thus affect the rotary diffusion process. However, as already mentioned by \citet{Hakansson_SAXSALIGN}, the value of this type of correction to $D_r$ is typically much lower than what is observed for CNF.

A final remark is that the fastest decay rates in this study were actually obtained in absence of extension. If the length distribution is truly the cause of the multi-time-scale decay, then it is maybe possible to capture a signature of the full length distribution by just utilizing a channel flow with high shear rate without any contraction. The experiment described here would thus be an easy characterization technique to estimate the length distribution of CNF in the dispersion.

\subsection{Other explanations}
The hypotheses of what is causing the experimental observations rely on the assumption that the flow really stops instantly and that the spatial distribution of fibrils is homogeneous over the channel cross section. 

First of all, the valves probably have a switching time of a few microseconds. It is not completely ruled out that there are no fluid dynamical phenomena occurring on this time scale when stopping the flow. This time scale can not be captured with the camera used in this experiment, and the flow does appear completely stopped with the frame rate used here. In order to completely rule out any motion of the flow on millisecond scale that could affect the fast rotary diffusion coefficient, the experiment must be analyzed thoroughly with high-speed cameras. 

Secondly, there could maybe be potential migration mechanisms involved in the system of CNF that result in local variations of the fibril concentration over the cross section. If there for example is a mechanism pushing fibrils towards the center of the channel, it will lead to a lower concentration of fibrils closer to the walls. This in turn might lead to a lower POM signal in the shear layers and potentially higher mobility of the fibrils. To analyze this further, the local concentration variations over the cross-section should be carefully examined.

Finally, when the dispersion is subject to shear, the resulting ODF will probably be slightly bi-axial. This means that the alignment will depend on the viewing angle in the plane perpendicular to the flow direction. It might also be possible that the diffusion process of this ODF will seem different depending on the viewing angle. In the converging channel geometry, the diffusion process in the shear layers is observed from different directions depending on if we look close to the lower/upper walls (viewing direction is along the vorticity direction) or if we look in the middle of the channel (viewing direction is along the gradient direction). Consequently, the bi-axiality of the ODF could possibly lead to difficulties in comparing the rotary diffusion process at different positions in the channel. To capture if the ODF really is bi-axial, the three-dimensional ODF at every spatial point in a channel cross section must be obtained, which possibly can be done with emerging tomographic SAXS techniques\cite{schaff2015six}.

\section{Conclusions}
In this work, we have presented a flow-stop experiment utilizing polarized optical microscopy (POM) for dynamic characterization of cellulose nanofibrils (CNF). The fibrils are aligned either through an extensional flow with low shear in a flow-focusing channel (FFC) geometry or through a converging channel (CC) geometry with varying combinations of shear and extension rates. By instantly stopping the flow, the decay of alignment can be measured and thus also the rotary diffusion coefficient $D_r$. By combining these experiments with SAXS measurements, the initial orientation distribution function (ODF) during steady state flow could also be measured. 

The main findings can be summarized with the following points:
\begin{itemize}
\item The decay of alignment after stopping the flow is not following a single relaxation time scale, but is rather a multiple time-scale process. The slowest decay, characterized by the rotary diffusion coefficient $D_{r,\text{slow}}$ remains almost constant, while the fastest decay, characterized by $D_{r,\text{fast}}$ is depending both on flow conditions and position in the channel.
\item Previous indications that $D_r$ of CNF would depend on the initial ODF are found to be an artefact of other causes since the same ODF can give two different values of $D_{r,\text{fast}}$.
\item The coefficient $D_{r,\text{fast}}$ is rather dependent on the local velocity gradients both in the streamwise direction (extension) and in cross-stream directions (shear). Higher extension rate $\dot{\varepsilon}$ and higher shear rate $\dot{\gamma}$  both result in a higher value of $D_{r,\text{fast}}$. High shear in the absence of extension gives a higher value of $D_{r,\text{fast}}$ than what is achieved in a pure extensional flow. Thus, a system of fibrils aligned with shear flow will de-align quicker than a system aligned by extensional flow.
\item The CC geometry allows for dynamical characterization at various average extension and shear rates ranging from pure extension with negligible shear (at the end of the contraction) to pure shear with negligible extension (downstream of the contraction). Different downstream positions in the contraction correspond to regions with competing effects from shear and extension.
\end{itemize}
There are several actual physical processes that could result in the observations summarised above. We hypothesize that various lengths of fibrils in the dispersion could be the cause of the different Brownian diffusion time scales and this would also explain the relationship between $D_{r,\text{fast}}$ and the strength of the velocity gradients. It is however not clear why there is a big difference in $D_{r,\text{fast}}$ between the cases of pure shear flow and pure extensional flow. We hypothesize that there could be resistive short-range interactions that are more likely to occur in the extensional flow and thus slowing down the rotary diffusion process. Since the initial ODF does not seem to influence $D_{r,\text{fast}}$, it is believed that electrostatic fibril-fibril interactions in the dispersion due to the surface charge of the fibrils do not play a major role in the rotary diffusion process.

In order to understand all the underlying processes, we propose future experiments using CNF dispersions with well defined fibril sizes and shapes and performing parametric variations of the concentration, the fibril charge, the pH of the solvent, the temperature or even adding different functional groups to the cellulose molecule.

Additionally, it would be interesting to use computational fluid dynamics (CFD) to determine the actual average shear rates $\dot{\gamma}$ and extension rates $\dot{\varepsilon}$  at the different positions in the channels. Combining experiments and simulations in this way could also increase our knowledge of how rotary diffusion of CNF is affected by velocity gradients. 

Even though the rotary diffusion process of CNF is not fully understood, the flow-stop experiment presented here is believed to be a valuable tool for dynamic characterization of CNF dispersions used for material production. In order to create a material with highly aligned fibrils, it would be important to try and minimize the fastest de-alignment time scales in the dispersion. The measurement with this device can consequently be used for a quick determination of the quality and usability of the nanocellulose prior to the material production process.

As a final remark, the presented flow-stop experiment is not limited to the study of cellulosic materials, but can be used to study any dispersion of non-spherical nanoparticles/macromolecules that become birefringent when aligned, e.g.~liquid crystals, proteins or polymers.

\section{Acknowledgments}
The authors acknowledge financial support from Wallenberg Wood Science Center (WWSC) and the Alf de Ruvo Memorial Foundation. A special thanks to Dr.~C.~Brouzet and Dr.~K.~H\aa kansson for the helpful discussions regarding both the POM experiments and the analysis of the SAXS results. Furthermore, the kind assistance during the SAXS experiments by Krishnegowda~V., Dr.~S.~Yu and Dr.~J.~McKenzie is gratefully acknowledged.

\pagebreak

\begin{center}
\section{Supplementary information: Dynamic characterization of cellulose nanofibrils in sheared and extended semi-dilute dispersions}
\end{center}
~\\

\subsection{POM flow-stop experiments}

The light illuminating the channel in the POM experiments can be decomposed into light polarized parallel and perpendicular to the $z$-direction. As the fibrils align in the $z$-direction, they also affect the birefringence of the dispersion. This means that light polarized in the $z$-direction will experience a different refractive index $n_\parallel$ when traveling through the dispersion than the refractive index $n_\perp$ for light with perpendicular polarization. The difference in refractive indices $\Delta n=n_\parallel-n_\perp$ causes a phase shift $\Delta \varphi$ between the two components. This phase shift can be measured by placing linear polarization filters on each side of the birefringent sample. These have polarization directions $45^{\circ}$ and $-45^{\circ}$ to the $z$-direction, respectively. The transmitted light intensity $I_{\text{POM}}$ through the filters and the sample is related to $\Delta n$ through the approximation\cite{Hakansson_RSC}:\\
\begin{equation}
I_{\text{POM}}\approx I_0\left(\frac{2\pi d}{\lambda}\right)^2(\Delta n)^2,
\end{equation}
~\\
where $d$ is the distance that the light has traveled through the sample, $\lambda$ is the wavelength of the light and $I_0$ is the intensity of the incoming light. This expression holds as long as $I\ll I_0$ and the angle that the sample has rotated the light is much less than $90^\circ$, which is assumed to be the case in the present experiments. Furthermore, the difference in refractive indices $\Delta n$ is related to the order parameter $S_\phi$ through \citep{vanGurp}:\\
\begin{equation}
S_\phi=\frac{\Delta n}{\Delta n_{\text{max}}},
\end{equation}
~\\
where $\Delta n_{\text{max}}$ corresponds to the difference in refractive index at perfect alignment ($S_\phi=1$). The order parameter can consequently be found in POM experiments using:\\
\begin{equation}
S_\phi=S_{\phi,\text{ref}}\sqrt{\frac{I_{\text{POM}}}{I_{\text{POM,ref}}}},
\label{eq:Paper9OrderVsI}
\end{equation}
~\\
given that an absolute value of the order parameter $S_{\phi,\text{ref}}$ and the corresponding light intensity $I_{\text{POM,ref}}$ are known for a certain reference case. Without these reference values, only relative measurements of the order parameter can be obtained.
At the same time, it is known that the order parameter of a dilute dispersion only subject to Brownian diffusion decays as\cite{lim1985conformation,rosenblatt1985birefringence,Hakansson_NatComm}\\
\begin{equation}
S_\phi(z,t)=S_{\phi,0}(z)\exp(-6D_rt).
\label{eq:diluteBrownian}
\end{equation}
~\\ 
Rotary diffusion coefficients can thus be obtained through the decay of the intensity signal $I_{\text{POM}}(t)$ by combining equations~\ref{eq:diluteBrownian} and \ref{eq:Paper9OrderVsI} and using the stationary order parameter at position $z$ as reference case:\\
\begin{equation}
I_{\text{POM}}(z,t)=I_{\text{POM},0}(z)e^{-12D_r t}.
\label{eq:Paper9DiffFromI}
\end{equation}
~\\
In order to only observe the influence of the light from the CNF, a background subtraction is done using measurements with water flowing through the channel. We also divide the subtracted intensity with the same background intensity, to remove any effects of the laser light not being completely uniform over the field of view, i.e.\\
\begin{equation}
I_{\text{POM,raw}}(z,t)=\frac{I_{\text{POM,CNF}}(z,t)-I_{\text{POM,background}}(z)}{I_{\text{POM,background}}(z)}.
\label{eq:Paper9IntensityAfterBG}
\end{equation}
~\\
The intensity in this case however does not decay to zero, as the optical properties of the equilibrium state of the CNF dispersion seems to be different from water. As this equilibrium level $I_{\text{eq.}}=I_{\text{POM,raw}}(z,t\rightarrow\infty)$ is not consistently positive or negative (after the background subtraction), we can not draw any conclusions about its origin. Therefore, the mean intensity during the last second of the experiment (around 10~s after stop) is set as the equilibrium level $I_{\text{eq.}}$ and the true intensity is given by $I_{\text{POM}}=I_{\text{POM,raw}}-I_{\text{eq.}}$. All negative values of the intensity after this subtraction are set to $10^{-16}$ in order to convert the intensity to an order parameter according to eq.~\ref{eq:Paper9OrderVsI}.

Each POM flow-stop experiment is initialized by starting the syringe pumps. After approximately 5~s of flow, the camera is recording 1500 frames with a rate of 100~frames per second. After approximately 5~s of recording the valves are switched to stop the flow in the flow cell, and the remaining time of approximately 10~s is recorded to measure the decay of birefringence. 

The steady state intensity during flow $I_{\text{POM,0}(z)}$ is taken as the mean intensity in a two second period, 3-5~s after the camera starts recording, just before the flow is stopped. The exact stopping time is found during post-processing as the time when the intensity drops below the standard deviation of the mean intensity.

\subsection{SAXS experiments}
\begin{figure}
\includegraphics[width=0.88\textwidth]{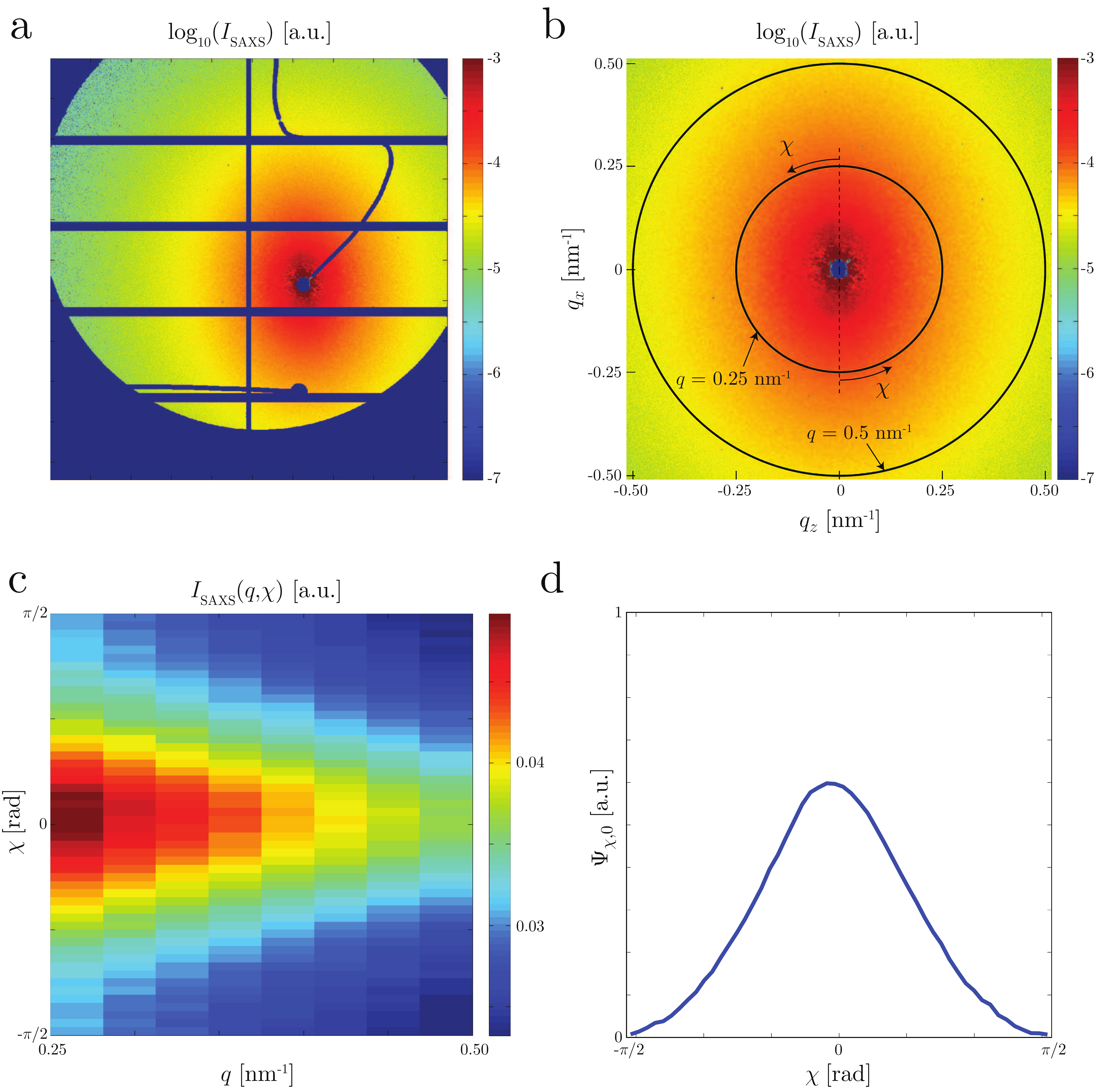}
\caption{\label{fig:Paper9SAXS} Example of results of the SAXS experiment at $z=2h$ in the flow-focusing setup; (a) original scattering image; (b) mirror corrected image; (c) histogram of scattering intensity $I_\text{SAXS}(q,\chi)$; (d) probability distribution $\Psi_{\chi,0}$.}
\end{figure}

As an X-ray beam is passing through the dispersion of CNF, the fluctuations of the electron distribution (due to the solid particles) gives rise to scattered light \citep{stribeck}. The scattered light intensity contains structural information of the particles and can therefore also be used to quantify the alignment of the CNF. The present procedure to analyze the SAXS data is similar to \citet{Hakansson_SAXSALIGN} and the additional considerations to obtain $S_\phi$ is described in detail by \citet{Rosen_MCSAXS}. 

The scattered light intensity is probed for different scattering vectors $\boldsymbol{q}$ with direction given by \citep{stribeck}:\\
\begin{equation}
\boldsymbol{e}_{\boldsymbol{q}}=\frac{\boldsymbol{e}_{\text{scat.}}-\boldsymbol{e}_{\text{inc.}}}{|\boldsymbol{e}_{\text{scat.}}-\boldsymbol{e}_{\text{inc.}}|},
\end{equation}
~\\
where $\boldsymbol{e}_{\text{scat.}}$ and $\boldsymbol{e}_{\text{inc.}}$ refer to unit vectors of the scattered and incident light, respectively. The length of the scattering vector is given by \citep{stribeck}:\\
\begin{equation}
|\boldsymbol{q}|=\frac{4\pi}{\lambda}\sin\varphi,
\end{equation}
~\\
where $\lambda$ is the wavelength of the X-ray and $2\varphi$ is the angle between $\boldsymbol{e}_{\text{scat.}}$ and $\boldsymbol{e}_{\text{inc.}}$. In small angle scattering experiments, the assumption $\sin\varphi\approx\varphi$ holds and we can probe the scattering intensity on a flat detector, where pixel coordinates $x_d$ and $z_d$ corresponds to scattering vector components \citep{stribeck}:\\
\begin{equation}
q_x=\frac{2\pi x_d}{\lambda D}, \qquad q_z=\frac{2\pi z_d}{\lambda D},
\label{eq:Paper9q}
\end{equation}
~\\
for the sample-to-detector distance $D$. To remove any contribution from the solvent, a background subtraction was done with scattering from pure water in the channel. An example of a scattering intensity image is showed in figure~\ref{fig:Paper9SAXS}a. The dead pixels on the detector and pixels behind the beamstop were first needed to be corrected. This was done by using the fact that the scattering pattern is symmetric and should ideally contain the same information at $(\pm q_x,\pm q_z)$. A pixel without information in the image was thus corrected by replacing its value by a value from any of the other three "mirror" pixels. The resulting intensity image after this procedure is illustrated in figure~\ref{fig:Paper9SAXS}b. The pixel coordinates are transformed to the cylindrical coordinates $q=\sqrt{q_x^2+q_z^2}$ and $\chi=\tan^{-1}(q_x/q_z)$. We then construct a histogram of $I_\text{SAXS}(q,\chi)$ in the range $q\in[0.25,0.5]$~nm$^{-1}$ (fig.~\ref{fig:Paper9SAXS}c) with certain number of bins in $q$- and $\chi$-directions. The number of bins in $q$- and $\chi$-direction was chosen to be significantly lower than the number of pixels in the $q$- and $\chi$-directions, respectively. This is ensuring that each histogram bin contains information. 

\begin{figure}
\includegraphics[width=0.4\textwidth]{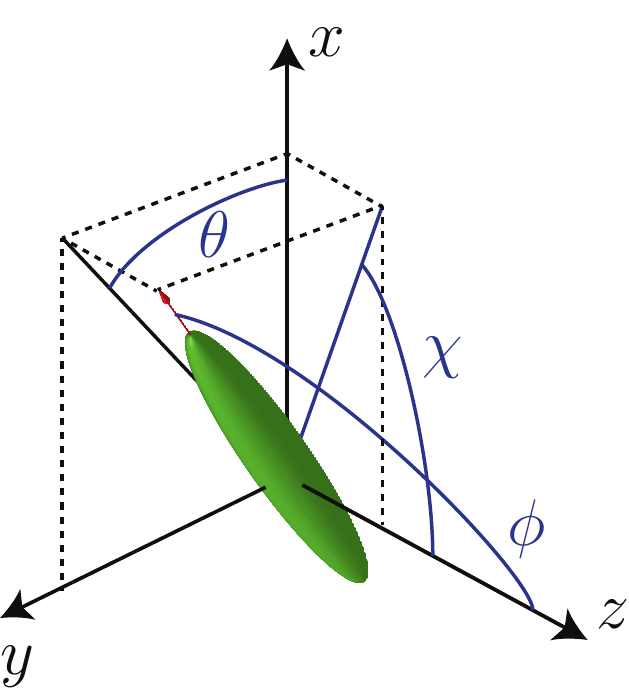}
\caption{\label{fig:ParticleOrientations} Orientation of an elongated particle described by the polar angle $\phi$ and the azimuthal angle $\theta$; the projected angle observed in the $xz$-plane is denoted $\chi$.}
\end{figure}

For each subrange of $q$, the isotropic contribution to the scattering is removed by assuming that the highest aligned case (at $z=1.5h$) has no fibers aligned perpendicular to the flow. This means that all intensity values in the $q$-subrange are subtracted by $I_{iso}(q)\approx I_\text{SAXS}(q,\chi=\pm\pi/2)$. Each subrange in $q$ is then normalized, the subranges are averaged and we obtain the steady state orientation distribution function (ODF) of $\Psi_{\chi,0}$ (fig.~\ref{fig:Paper9SAXS}d). The measurement thus gives the average projected angle $\chi$ of particles in the viewing ($xz$) plane, as illustrated in figure~\ref{fig:ParticleOrientations}. This angle is related to $\phi$ and $\theta$ through the relationship:\\
\begin{equation}
\tan\chi=\tan\phi\cos\theta.
\end{equation}
~\\
To obtain the ODF of the polar angle $\phi$, we make the assumption that the azimuthal angle $\theta$ is distributed uniformly in the flow. The steady state ODF $\Psi_{\phi,0}$ is then obtained with the same method as described by \citet{Rosen_MCSAXS}. The corresponding steady state order parameter $S_{\phi,0}$ is found through:
\begin{equation}
S_{\phi,0}=\int_{-\pi/2}^{\pi/2}\Psi_{\phi,0}\left(\frac{3}{2}\cos^2\phi-\frac{1}{2}\right)|\sin\phi|d\phi.
\end{equation} 
~\\
with normalization
\begin{equation}
\int_{-\pi/2}^{\pi/2}\Psi_\phi|\sin\phi|d\phi=1.
\end{equation}

\subsection{Estimating the average shear rate in a fully developed channel flow}
The velocity $w$ at a radial position $r$ of a fully developed flow in a circular pipe with radius $R$ and streamwise pressure gradient $dp/dz$ is given directly from the incompressible Navier-Stokes equations as:\\
\begin{equation}
w(r)=\frac{1}{4\mu}\frac{dp}{dz}(R^2-r^2),
\end{equation}
~\\
where $\mu$ is the dynamic viscosity of the fluid. The flow rate $Q$ can be related to the pressure gradient through:\\
\begin{equation}
Q=\int_0^{2\pi}\int_0^R w(r)rdrd\phi=\frac{2\pi}{4\mu}\frac{dp}{dz}\int_0^R (rR^2-r^3)dr=\frac{\pi}{2\mu}\frac{dp}{dz}\left[\frac{r^2R^2}{2}-\frac{r^4}{4}\right]_0^R=\frac{\pi R^4}{8\mu}\frac{dp}{dz}
\label{eq:FlowRatePressGrad}
\end{equation}
~\\
The average shear rate $\dot{\gamma}_\text{avg.,pipe}$ in the pipe will be given by:\\
\begin{equation}
\dot{\gamma}_\text{avg.,pipe}=\frac{1}{\pi R^2}\int_0^{2\pi}\int_0^R\left|\frac{dw}{dr}\right|rdrd\phi=\frac{2}{ R^2}\int_0^R\frac{1}{2\mu}\frac{dp}{dz}r^2dr=\frac{R}{3\mu}\frac{dp}{dz}.
\end{equation}
~\\
Using the result in eq.~(\ref{eq:FlowRatePressGrad}), this can be rewritten to:\\
\begin{equation}
\dot{\gamma}_\text{avg.,pipe}=\frac{R}{3\mu}\frac{8\mu Q}{\pi R^4}=\frac{8Q}{3\pi R^3}.
\end{equation}
~\\
Assuming the same average shear rate in a quadratic channel with side $h$ and same cross-sectional area, i.e.~$R=h/\sqrt{\pi}$, we finally obtain:\\
\begin{equation}
\dot{\gamma}_\text{avg.}=\frac{8Q}{3\pi (\frac{h}{\sqrt{\pi}})^3}=\frac{8Q\sqrt{\pi}}{3h^3}.
\end{equation}
~\\

\subsection{The two-dimensional Smoluchowski equation}

The steady state two-dimensional ODF $\Psi_{\phi_\text{2D}}$ is given by the solution of the stationary 2D Smoluchowski equation\cite{doi1986theory}:~\\
\begin{equation}
\frac{\partial^2\Psi_{\phi_\text{2D}}}{\partial\phi_\text{2D}^2}-Pe\frac{\partial(\Psi_{\phi_\text{2D}}\dot{\phi}_\text{2D})}{\partial\phi_\text{2D}}=0.
\end{equation}
~\\
In the simulations, the particles are assumed to be prolate spheroids with aspect ratio (length/width) $r_p=100$. 

In a shear flow, the P\'eclet number is given by $Pe=\dot{\gamma}/D_r$, where $\dot{\gamma}$ is the shear rate and $D_r$ is the rotary diffusion coefficient. The angular velocity $\dot{\phi}_\text{2D}$ is given by \citet{Jeffery} as:~\\
\begin{equation}
\dot{\phi}_\text{2D}=\frac{r_p^2-1}{r_p^2+1}\left(\frac{1}{2} - \sin^2 \phi_\text{2D}\right)-\frac{1}{2}.
\end{equation}
~\\
In a planar extensional flow, the P\'eclet number is given by $Pe=\dot{\varepsilon}/D_r$, where $\dot{\varepsilon}$ is the extension rate. In this case the angular velocity is given by \cite{Jeffery}:\\
\begin{equation}
\dot{\phi}_\text{2D}=-\frac{r_p^2-1}{r_p^2+1} 2 \cos\phi_\text{2D}\sin\phi_\text{2D}.
\end{equation}
\bibliography{REFERENCES}

\end{document}